# Embedded quantum computing for many-body surface reaction

Dedong Wan[1,2], Xiaopeng Li[1], Yi Fan[1], Jie Liu[3], Xiongzhi Zeng[1]*, Zhenyu Li[1]*

[1]State Key Laboratory of Precision and Intelligent Chemistry, University of Science and Technology of China, Hefei 230026, China

[2]Hefei National Research Center for Physical Sciences at the Microscale, University of Science and Technology of China, Hefei, 230026, China

[3] Hefei National Laboratory, University of Science and Technology of China, Hefei 230088, China

*To whom correspondence should be addressed:

xzzeng@ustc.edu.cn; zyli@ustc.edu.cn

## Abstract

We introduce QC-DFET, a quantum-computing-based density-functional embedding framework for accurately simulating reactions at catalytic interfaces. Within the embedded reactive region, a reaction-consistent protocol selects active orbitals and maintains orbital continuity along reaction coordinates. The selected orbitals define compact active-space Hamiltonians that incorporate the influence of the surrounding surface and are subsequently mapped onto qubits. Quantum-selected configuration interaction combines quantum sampling on the *Zuchongzhi* superconducting quantum processor, using up to 40 qubits, with classical diagonalization to describe electron correlation within these active spaces. Strongly contracted perturbation theory then accounts for dynamic correlation beyond the active space. We validate QC-DFET through two experimentally constrained surface-chemistry challenges on Cu(111) and Cu(100): $H_2$ dissociation/desorption tests balanced bond breaking and recombination barriers, while CO adsorption tests site selectivity and metal–adsorbate bonding on both surfaces. Across these cases, QC-DFET reproduces bidirectional $H_2$ barriers and recovers the observed top-site preference and adsorption strength of CO on both Cu surfaces. These results establish embedded quantum computing as a practical route to correlated surface-reaction energetics.

Surface reactions control the activity and selectivity of heterogeneous catalysts through elementary bond-breaking and bond-forming events at extended interfaces.[1-4] Predictive simulations of these events require more than locating stationary points on a potential-energy surface: the computed energetics must balance adsorbate bond rearrangement, metal-adsorbate charge redistribution and the electronic response of the surrounding surface. This balance is particularly demanding on transition-metal catalysts, where localized adsorbate frontier orbitals hybridize with delocalized metallic states. Errors of a few tenths of an electron-volt can change the predicted adsorption site, reverse the ordering of competing reaction channels or obscure the connection between atomistic mechanisms and experimental observables.

These competing requirements have made density functional theory (DFT) the workhorse of surface chemistry, while also exposing the approximations that limit most practical surface calculations. Semilocal and dispersion-corrected generalized-gradient approximations make periodic slab simulations tractable at routine cost; however, self-interaction, delocalization and density-driven errors can distort metal–adsorbate charge redistribution and lead to unreliable reaction and adsorption energetics.[5–7] Such errors are not merely technical: they are manifested in benchmark problems such as $H_2$ dissociation on Cu(111) and CO adsorption on transition-metal surfaces, where barrier heights and site preferences depend sensitively on electron correlation and surface charge response.[4,8–21] Double-hybrid density functionals, random-phase approximation, quantum Monte Carlo and embedded correlated-wavefunction methods have improved selected cases, but their computational cost and system dependence still limit routine application to complex catalytic interfaces.[14-17,21–23]

Wave-function-based electronic-structure methods can systematically improve the treatment of electron correlation, but their direct application to extended surfaces is limited by the steep scaling with the number of correlated electrons and orbitals.[24-26] Embedding approaches address this difficulty by treating a chemically important region at high level while retaining the surrounding environment at lower cost.[27-29] In density-functional embedding theory (DFET), a cluster containing the reactive site is embedded in an effective potential derived from the full periodic system, allowing correlated calculations to include the effect of the extended surface.[30-33] DFET-based embedded correlated-wavefunction calculations have clarified surface-reaction problems including $H_2$ desorption from Cu(111), methane activation and electrochemical $CO_2$-reduction intermediates.[34-35] These studies show that the essential many-body corrections in surface chemistry are often local enough to be embedded, but the active-space solver remains a central bottleneck.

Quantum computing provides a promising route to this bottleneck by representing many-electron active-space wave functions in qubit Hilbert spaces.[36,37] Variational quantum eigensolvers, adaptive ansatz construction and fermionic excitation circuits have enabled increasingly compact quantum simulations of molecular electronic structure.[38-45] More recently, quantum-selected configuration interaction (QSCI) and related quantum-centric diagonalization methods have shifted the near-term role of quantum hardware from full variational optimization to the sampling of important determinants, followed by classical diagonalization in the selected subspace.[46-48] This strategy is attractive for chemistry because the quantum device is used for a naturally probabilistic task—identifying important configurations—while the final active-space energy remains variational within the selected determinant space.

Applying this idea to catalytic surfaces requires more than a quantum active-space solver. The active-space Hamiltonian must be small enough for present quantum hardware, chemically complete enough to describe bond rearrangement and physically embedded enough to retain the metallic environment. Earlier quantum-computing studies of surface reactions established local-embedding workflows and demonstrated feasibility for surface models, but often remained at the proof-of-concept level or used simplified clusters.[49-50] In parallel, recent many-body embedding and screening approaches for metal surfaces have demonstrated that high-level solvers can recover experimentally constrained adsorption and reaction energetics.[19,22-23] These advances point to an interesting direction: incorporate quantum computing into an embedding framework to treat realistic, experimentally benchmarked surface reactions rather than isolated molecular or cluster models.

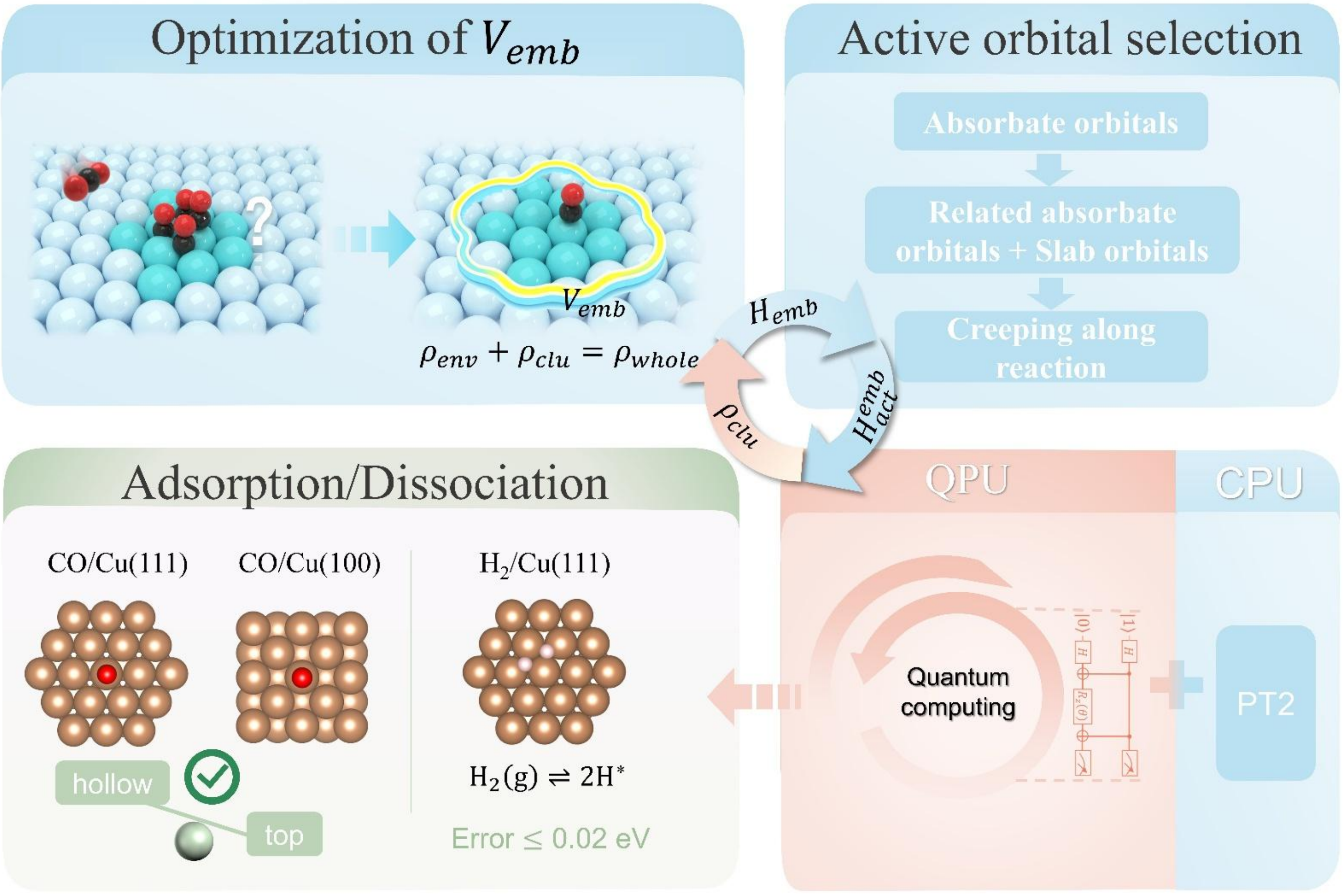

**Fig. 1** Workflow of the QC-DFET implementation. Classical-computing components are shown in blue, quantum-hardware components are highlighted in pink beige, and the resulting QC-DFET outputs are marked in green.

Along this direction, in this study, we introduce quantum-computing density-functional embedding theory, QC-DFET, for correlated surface-reaction energetics. QC-DFET constructs DFET-derived active-space Hamiltonians that retain the influence of the extended metallic environment and maps them onto qubits, while concentrating quantum resources on the adsorbate-metal orbitals responsible for bond rearrangement, charge redistribution and site-specific bonding. The many-body expansion-based complete active space for surface reaction (MBECAS-SR) protocol selects these orbitals and maintains orbital consistency along reaction coordinates, yielding compact active spaces for quantum computation. QSCI is then used to solve these Hamiltonians on the *Zuchongzhi* superconducting quantum processor, while residual dynamic correlation is recovered through strongly contracted NEVPT2-based corrections. We demonstrate the framework using experimentally constrained benchmarks on Cu surfaces that test complementary requirements for predictive surface chemistry. $H_2$ dissociation/desorption on Cu(111) inspects whether QC-DFET can provide balanced forward and reverse barriers for coupled H–H bond cleavage and H–Cu bond formation. CO adsorption on Cu(111) and Cu(100) tests whether the method can recover adsorption strengths and site preferences while providing an orbital-resolved picture of Cu–CO bonding. Across these applications, QC-DFET brings quantum-hardware sampling into environment-aware many-body simulations of catalytic surfaces and provides a practical computational route for experimentally constrained surface-reaction energetics.

## Results

### QC-DFET workflow for experimentally benchmarked surface reactions

The QC-DFET architecture separates the surface-reaction problem into three linked tasks: preserving the metallic environment, choosing an active space that remains chemically continuous across relevant structures, and extracting quantitatively useful many-body energies from near-term quantum states. We benchmark QC-DFET through three Cu-surface challenges: state-balanced $H_2$ bond activation on Cu(111) and site-selective CO adsorption on Cu(111) and Cu(100).

**$H_2$ dissociation/desorption.** We first examine $H_2$ dissociation and the reverse associative desorption of adsorbed H atoms on Cu(111), a stringent benchmark for state-balanced surface-reaction energetics. Along this single reaction coordinate, the electronic structure evolves from two Cu-bound H fragments in the reactant, through a strongly coupled transition state, to molecular H-H bonding in the product. Accurate barriers therefore require matched errors for the initial, transition and final states, not merely an accurate description of one endpoint. The process is experimentally constrained in both directions: the dissociative adsorption barrier is 0.77 eV, whereas the zero-point-energy-corrected associative desorption barrier is 0.59 eV.[8,9] These paired reference values provide a direct test of whether one embedded active-space description remains comparably accurate across all three states.

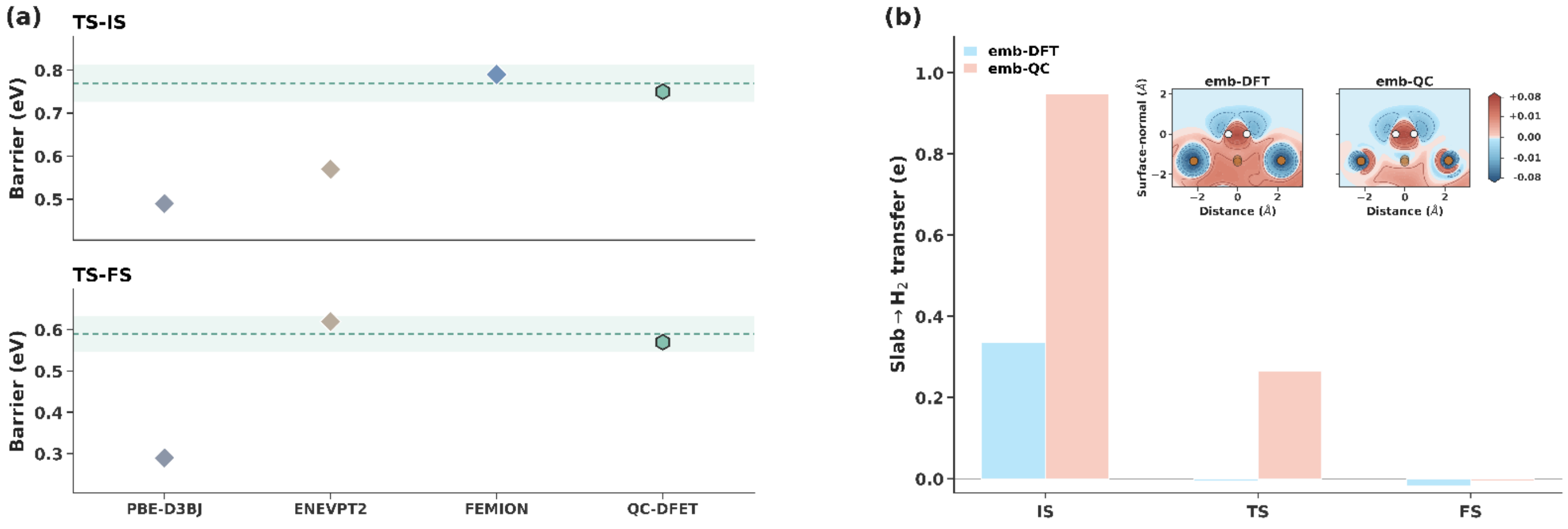

**Fig. 2 Benchmark validation of QC-DFET for $H_2$ dissociation/desorption on Cu(111).** (a) Calculated TS-IS and TS-FS barriers (IS, TS, and FS denote the initial, transition, and final states, respectively) using PBE-D3BJ, ENEVPT2[34], FEMION[23] and QC-DFET. Dashed lines indicate experimental barriers; shaded bands denote chemical accuracy (0.043 eV). (b) Net Mulliken charge transfer from the Cu slab to the H adsorbate for IS, TS, and FS, evaluated at the emb-DFT and emb-QC levels. Positive values denote net electron donation from the slab to the adsorbate. The inset compares the real-space electron-density rearrangements at the TS relative to a common neutral-atom promolecular reference (in e bohr$^{-3}$). Red and blue regions indicate electron-density accumulation and depletion relative to this reference, respectively.

The $H_2$/Cu(111) reaction geometries were optimized at the PBE-D3BJ level using a $\sqrt{21} \times \sqrt{21}$ four-layer Cu(111) slab.[34] For the embedded calculations, the $Cu_{12}$ cluster and the corresponding embedding potential were directly adopted from the previous ENEVPT2 study[34], enabling a consistent comparison with the reported embedded-wavefunction results. The cluster contains the first-layer Cu atoms directly involved in the reaction together with a small number of nearby second-layer Cu atoms, while the total electron number of the cluster–adsorbate system is kept even to avoid an artificial odd-electron cluster. The resulting embedded model and embedding potential are shown in Fig. S1.

Using this embedded model, we evaluate the TS–IS and TS–FS barriers (IS, TS, and FS denote the initial, transition, and final states, respectively) and compare the QC-DFET results with previous theoretical predictions and experimental references (Fig. 2a). PBE-D3BJ underestimates both the TS-IS and TS-FS barriers by 0.28 and 0.30 eV, respectively, indicating that its errors in the initial, transition and final states are not balanced for this reaction coordinate. The DFET-based ENEVPT2 treatment[34] improves the reverse barrier but still underestimates the forward barrier by 0.20 eV, suggesting that the active-space description is less consistent between the transition state and the dissociated-H state. FEMION[23] predicts a barrier of 0.79 eV, in close agreement with experiment. In contrast, QC-DFET, using a CAS(12e,12o) embedded active-space Hamiltonian mapped to 24 qubits, predicts both the TS–IS and TS–FS barriers with absolute errors of only 0.02 eV, respectively. This agreement shows that QC-DFET reduces state-dependent error imbalance across the reactant, transition state and product.

This state balance arises from the reaction-consistent construction of the active space. In the QC-DFET active space, two surface-coupled adsorbate-centered orbital pairs are retained for the separated H atoms (Fig. S4a), allowing the active-space manifold to evolve continuously from the two Cu-bound H fragments to an H-H bonding/antibonding description. This continuity is essential because the orbitals that dominate the initial, transition and final states change character as the H-H bond breaks and H-Cu interactions form, otherwise a mismatch in accuracy of forward and backward barriers is expected.[34] The $H_2$ benchmark shows that the accuracy of QC-DFET depends not only on the correlated solver, but also on maintaining a chemically continuous active space that gives matched-quality descriptions of all stationary states along the reaction coordinate.

Electron correlation also modifies the adsorbate–surface electronic response along the reaction coordinate. To investigate such responses, we check the results of embedded quantum-computing (emb-QC) and embedded DFT (emb-DFT) cluster calculations in the presence of the embedding potential. As shown in Fig. 2b, the net slab-to-adsorbate charge transfer nearly vanishes at the TS at the emb-DFT level, whereas emb-QC retains appreciable electron donation from the Cu slab to the H adsorbate. Interestingly, this difference is associated with the delocalization error of semilocal DFT. The inset of Fig. 2b further compares the real-space electron-density rearrangements at the TS relative to a common neutral-atom promolecular reference (Supplementary Section 3.1). Both methods show electron-density accumulation around the H adsorbate and depletion in neighboring regions. However, emb-DFT exhibits a more continuous accumulation region extending toward the Cu surface, whereas emb-QC shows a clearer spatial separation between the accumulation near the H atoms and the density redistribution around the surface Cu atoms. This picture is consistent with fractional-electron energies (Fig. S6 and Supplementary Section 3.2).

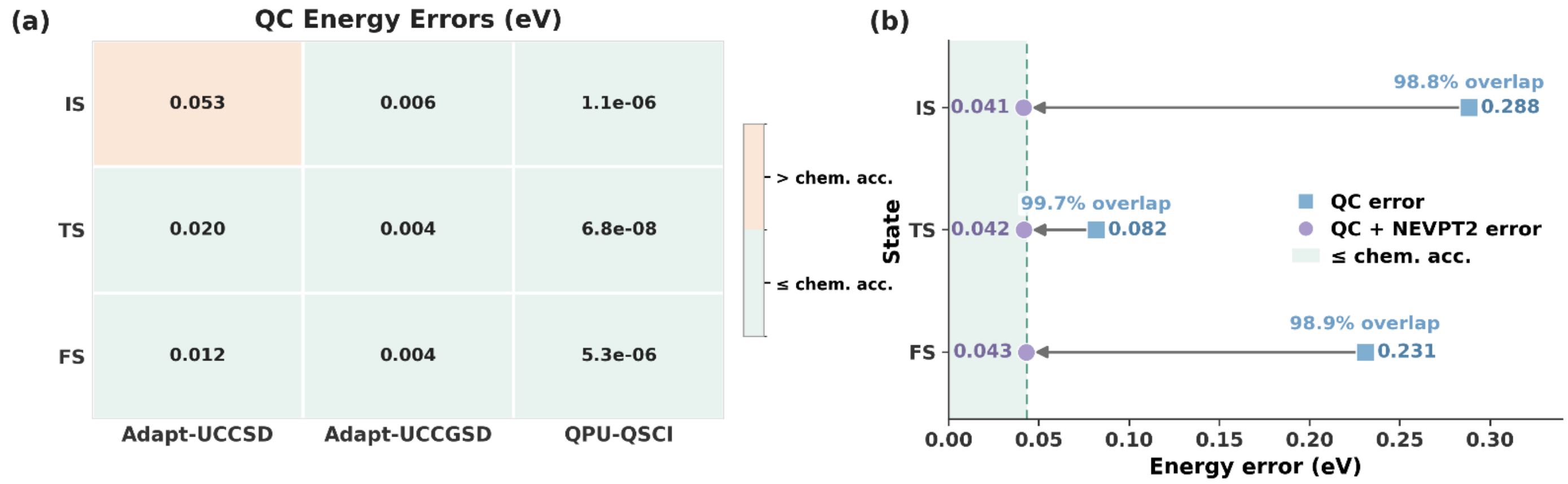

**Fig. 3 Quantum-computing energy errors and NEVPT2-assisted error reduction for $H_2$ dissociation/desorption on Cu (111).** (a) State-resolved QC energy errors obtained with ADAPT-UCCSD, ADAPT-UCCGSD and QPU-QSCI. The QPU-QSCI results correspond to the recovered QSCI configuration space obtained from the hardware. Cell colors indicate whether the corresponding energy error is within chemical accuracy. (b) NEVPT2 correction reduces raw QC energy errors for all three states to within chemical accuracy. Blue squares denote raw QC errors, purple circles denote QC+NEVPT2 errors, arrows indicate error reduction and the shaded region marks chemical accuracy. Percentages above QC-error points give the corresponding wave function overlaps with FCI.

Quantum computing is central to QC-DFET because it supplies the correlated active-space states from which the embedded IS, TS and FS energies are constructed. We therefore examine how the choice of quantum-computing active-space solver affects the state-resolved description of the $H_2$ reaction coordinate. Figure 3a compares active-space energy errors for the IS, TS and FS obtained from ADAPT-VQE using unitary coupled-cluster singles and doubles (UCCSD) and generalized singles and doubles (UCCGSD) operator pools, together with QSCI. ADAPT-UCCSD and ADAPT-UCCGSD results are obtained via noiseless simulations. ADAPT-UCCSD gives small but not uniformly chemically accurate errors across the three states, whereas the more flexible ADAPT-UCCGSD pool reduces all three errors to below 0.01 eV. Thus, the active-space solver must be sufficiently expressive to deliver matched-quality descriptions of the reactant, transition state and product. Importantly, QPU-QSCI reaches a comparable level of active-space accuracy for all three geometries despite quantum-device noise, showing that determinant subspaces recovered from measured bitstrings are already sufficiently informative for reconstructing the relevant embedded active-space energies.

Then, we consider perturbative corrections that recover dynamic correlation beyond the quantum-computed active space.[45] Specifically, we assess the sensitivity of SC-NEVPT2-corrected energies to incomplete convergence of the active-space wave functions by applying SC-NEVPT2 to a series of intermediate states sampled along the ADAPT-UCCGSD convergence trajectories (Fig. 3b and Fig. S10). The corrected total-energy errors remain within chemical accuracy even for intermediate states with raw active-space QC errors of approximately 0.08 eV for TS and 0.23–0.29 eV for IS and FS. Considering that the overlaps between the intermediate and FCI wave functions are high, we can conclude that the perturbative correction can tolerate imperfect active-space states once the dominant correlation pattern is captured. Thus, QC-DFET can benefit from two complementary requirements: a sufficiently expressive quantum solver improves the active-space state itself, and SC-NEVPT2 reduces the impact of residual state-dependent solver errors.

**CO adsorption-site puzzle.** We then applied QC-DFET to CO adsorption on Cu(111), a prototypical adsorption-site puzzle in which commonly used GGA functionals over-stabilize high-coordination hollow sites relative to the experimentally observed top site.[10,11] This problem is demanding because the site preference depends on a subtle balance among CO 5σ donation, metal-to-CO 2π* back-donation, direct C-Cu bonding and surface charge redistribution.[18] It therefore tests two aspects of QC-DFET: whether the method can recover the adsorption-energy ordering, and whether the correlated embedded description provides a chemically interpretable picture of the bonding channels behind that ordering.

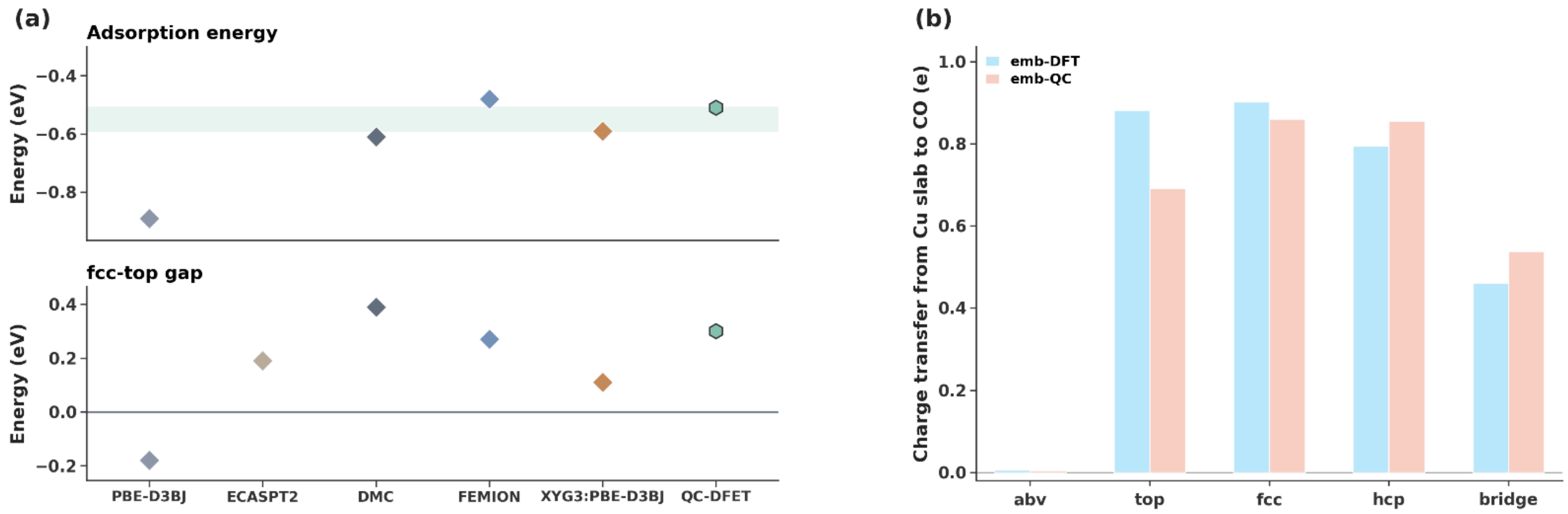

**Fig. 4 Adsorption energetics and charge-transfer analysis for CO on Cu (111).** (a) Comparison of CO adsorption energetics from PBE-D3BJ, ECASPT2[15], DMC[16], FEMION[23], XYG3:PBE-D3BJ[17] and QC-DFET. The upper panel shows the adsorption energy at the top site, while the lower panel shows the fcc-top energy difference. The shaded region in the adsorption-energy panel denotes the experimental reference window centered at -0.55 eV[12] with chemical-accuracy width. The solid zero line in the fcc-top panel separates predictions favoring top and fcc sites. (b) Mulliken charge transfer from the Cu slab to CO for different adsorption configurations at the emb-DFT and emb-QC levels. Positive values correspond to net electron donation from the metallic substrate to CO.

Figure 4a summarizes the adsorption energetics. PBE-D3BJ overestimates the CO adsorption strength and incorrectly favors the fcc hollow site over the top site. Higher-level approaches, including DFET-based ECASPT2, DMC, FEMION and XYG3, correct this qualitative error and recover the top-site preference.[15-17,23] QC-DFET, using QSCI for a CAS(14e,12o) active-space Hamiltonian derived from an embedded $Cu_{14}$ cluster carved from a four-layer 5 × 5 Cu(111) slab and mapped to 24 qubits, also predicts the top site as the most stable adsorption configuration. The top-site adsorption energy lies within the chemical-accuracy window of the experimental reference[12], and the fcc-top energy separation is 0.30 eV, close to the DMC value and larger than the gaps predicted by several other high-level methods. Across the four adsorption sites, QC-DFET gives the stability order top > bridge > fcc > hcp (Fig. S13). XYG3 also predicts the same ordering but with a weaker top-site preference. However, the relative energy differences among these adsorption sites lack quantitative experimental benchmarks and require further validation.

We also tested the sensitivity of the QC-DFET energetics to embedded-cluster size. Calculations with the smaller $Cu_{12}$ cluster used in the previous ECASPT2 work[15] still give a top-site adsorption energy close to experiment and retain a substantial fcc–top separation (Table S1). However, the hcp site is affected by a finite-cluster artifact because one selected slab orbital has pronounced Cu 3d edge character[15]. These results indicate that the main top/fcc energetics are robust, whereas a complete description of the full multi-site energy landscape requires embedded clusters large enough to avoid edge-localized orbital artifacts.

The corrected adsorption energetics motivate a more detailed analysis of the CO–Cu electronic structure. The Mulliken charge-transfer analysis in Fig. 4b shows that both emb-DFT and emb-QC predict similar electron donation from Cu to CO for the chemisorbed configurations, indicating that the recovered top-site preference cannot be explained by charge transfer alone. A more local, orbital-resolved bonding descriptor is therefore needed.

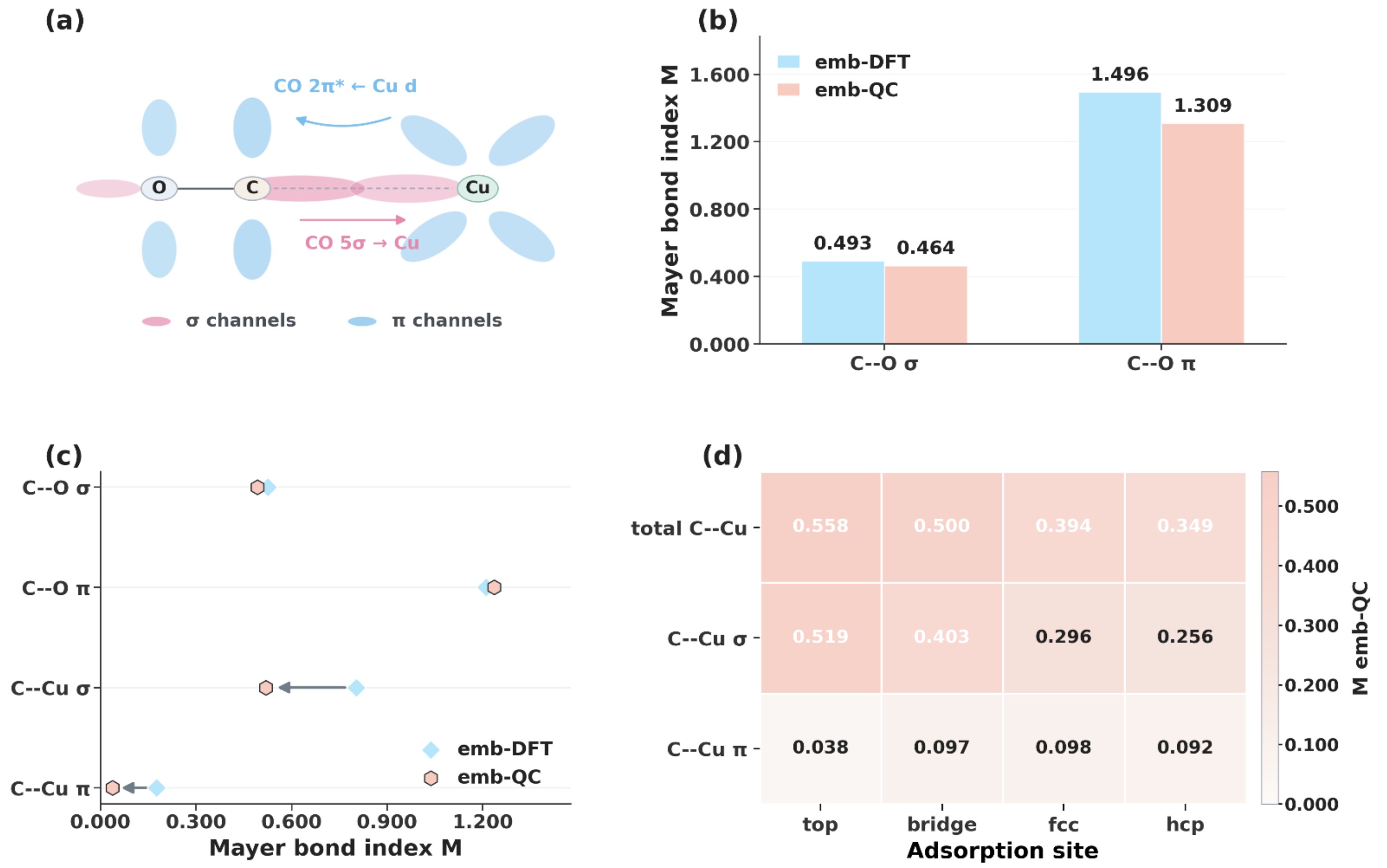

**Fig. 5 Orbital-resolved Mayer bond-order analysis of CO adsorption on Cu (111).** (a) Qualitative schematic of the orbital interactions underlying the Mayer bond-order channels, showing CO 5σ donation along the Cu–C axis and π back-donation from Cu d states into the CO 2π* manifold. (b) Mayer bond indices of the internal C-O σ and π components for the free-CO reference configuration, comparing emb-DFT and emb-QC descriptions. (c) Top-site orbital-resolved Mayer bond indices from emb-DFT and emb-QC for internal C-O σ/π components and C-Cu σ/π coupling channels. Arrows highlight the DFT-to-QC correction for direct C-Cu coupling channels. (d) Site-resolved emb-QC C-Cu coupling on Cu(111), decomposed into total, σ and π components. The top site exhibits the largest total C-Cu coupling, dominated by the σ component, followed by bridge, fcc and hcp sites, consistent with the QC-DFET adsorption-site stability trend.

To resolve these bonding channels, we performed orbital-resolved Mayer bond-order analysis based on the Blyholder picture of CO-metal bonding (Fig. 5a).[18] The free-CO reference establishes how the correlated embedded treatment modifies the internal CO bond in the absence of C-Cu coupling: emb-QC and emb-DFT give similar internal C-O σ and π Mayer indices (Fig. 5b). For top-site adsorption, the internal C-O σ and π components still change only weakly from emb-DFT to emb-QC, however, the direct C-Cu coupling is substantially reduced, mainly through the σ channel in absolute magnitude (Fig. 5c). Thus, the dominant DFT-to-QC correction in the adsorbed state is not a restructuring of the internal CO bond, but a correction to the localized metal-adsorbate coupling.

The site-resolved emb-QC Mayer indices then connect this bonding correction to the QC-DFET energy landscape. The total C-Cu coupling is largest at the top site and decreases through bridge, fcc and hcp sites, following the QC-DFET stability order. Decomposition into σ and π components shows that this ordering is dominated by localized C-Cu σ coupling, whereas the π contribution is smaller and does not compensate for the weaker direct σ interaction at bridge and hollow sites. In summary, the physical picture is therefore that QC-DFET recovers the top-site preference not by maximizing net slab-to-CO charge transfer, but by correcting the local orbital channels so that the strongest C-Cu σ coupling coincides with the most stable adsorption site.

**40-qubit CO model.** Having established the adsorption-site energetics and their bonding origin on Cu(111), we further extend QC-DFET to CO adsorption on Cu(100) using hardware-sampled QSCI with a 40-qubit Hamiltonian derived from an embedded $Cu_{14}$ cluster carved from a four-layer 5 × 5 Cu(100) slab. For this larger quantum calculation, we introduce correlation-directed self-consistent configuration recovery (CD-SCORE), complemented by Hamiltonian-guided[52] refinement, and employ the uncontracted Epstein–Nesbet perturbation theory approximation to SC-NEVPT2 based on spin-pure selected-CI wave functions (EN-SC-NEVPT2)[54] to incorporate dynamic correlation at a tractable computational cost (Supplementary Sections 5.3 and 5.4).

Figure 6a compares the calculated top-site CO adsorption energies with the experimental reference[13]. PBE-D3BJ, PW91[20], and SCAN[14] substantially overestimate the adsorption strength, whereas RPA[14] gives an adsorption energy close to experiment[13] and dispersion-corrected emb-MC-ftPBE[21] predicts somewhat weaker adsorption. QC-DFET yields a top-site adsorption energy of approximately −0.59 eV, in close agreement with the experimental value. Importantly, QC-DFET also correctly identifies the top site as the most stable adsorption configuration, with bridge–top and hollow–top energy separations of approximately 0.36 and 0.76 eV, respectively (Fig. 6b). In contrast, PBE-D3BJ, PW91, and SCAN slightly favor higher-coordination sites, while RPA and emb-MC-ftPBE also recover the top-site preference with smaller energy separations.

The orbital-resolved Mayer bond-order analysis on Cu(100) further examines the bonding mechanism underlying the adsorption-site preference. The top-site DFT-to-QC changes closely resemble those on Cu(111) (Fig. 6c). Across the Cu(100) adsorption sites, the C–Cu σ component decreases in the order top > bridge > hollow, consistent with the calculated site stability. By contrast, the π contribution increases at higher-coordination sites, consistent with additional lateral coupling channels between C 2p and neighboring Cu 3d orbitals. At the bridge site, this increase nearly compensates for the modest reduction in σ coupling, yielding a total C–Cu Mayer bond index comparable to that at the top site (Fig. 6d). The Cu(100) site preference therefore correlates more directly with the localized C–Cu σ component than with the total C–Cu bond index.

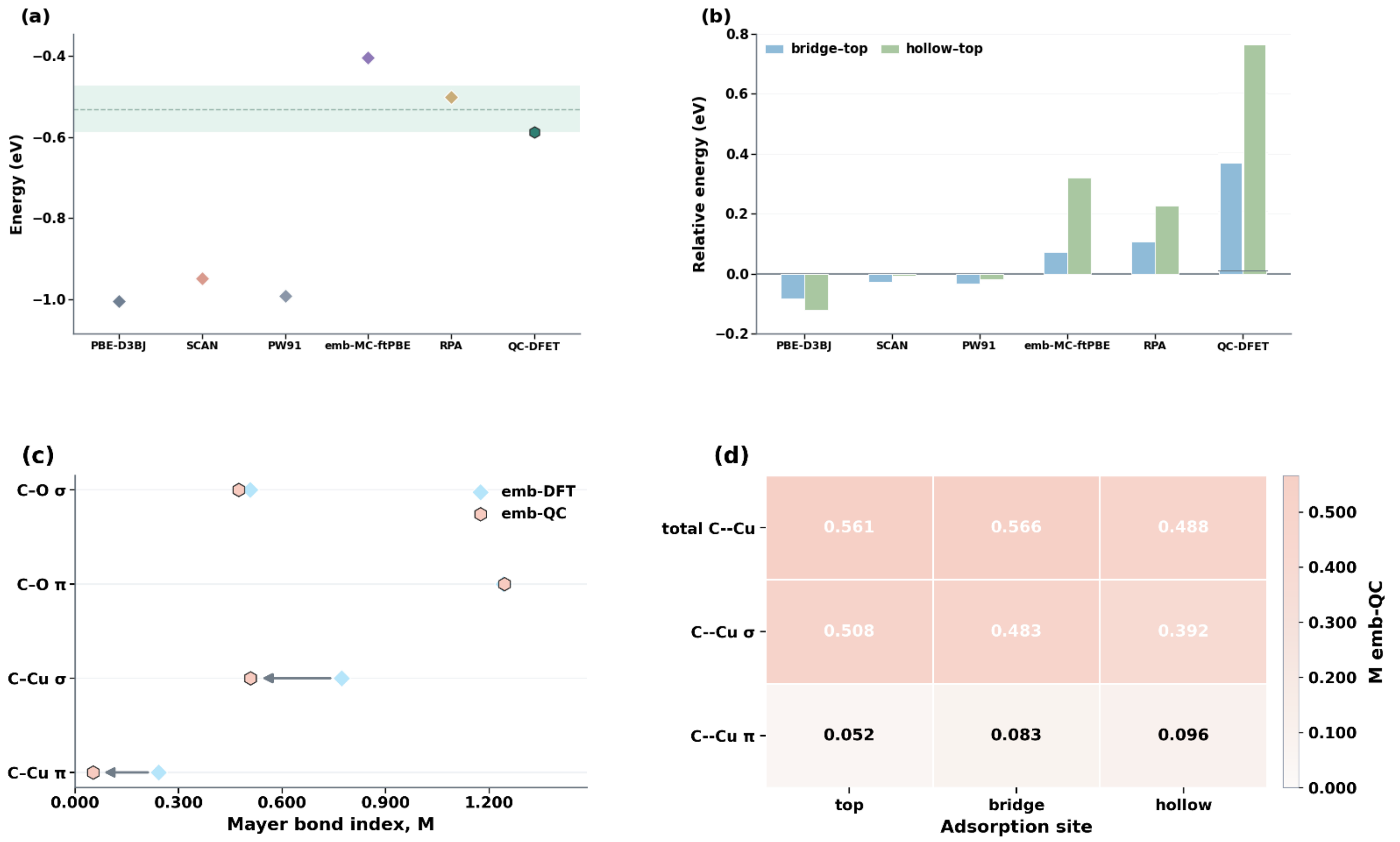

**Fig. 6 CO adsorption energetics and orbital-resolved bonding analysis on Cu(100).** (a) Top-site CO adsorption energies calculated using PBE-D3BJ, SCAN[14], PW91[20], dispersion-corrected emb-MC-ftPBE[21], RPA[14], and 40-qubit QC-DFET. The shaded band represents the low-coverage experimental adsorption-energy benchmark from TDS measurements (Ref. 13), including the reported experimental uncertainty and an additional chemical-accuracy tolerance. (b) Bridge–top and hollow–top relative adsorption energies obtained using the same methods. Positive values indicate a preference for the top site, whereas negative values favor higher-coordination sites. (c) Orbital-resolved Mayer bond indices for top-site CO adsorption, comparing the emb-DFT and emb-QC descriptions of the internal C–O σ/π components and direct C–Cu σ/π coupling channels. Arrows highlight the DFT-to-QC changes in the C–Cu coupling channels. (d) Site-resolved C–Cu Mayer bond indices from the emb-QC calculations, decomposed into total, σ, and π components for the top, bridge, and hollow sites.

## Discussion

QC-DFET establishes an embedded quantum-computing strategy for correlated surface-reaction energetics. The framework addresses three requirements that must be satisfied simultaneously before quantum computing can become useful for realistic catalytic interfaces. First, the active-space Hamiltonian must retain the effect of the extended surface environment; here this is achieved through a DFET-derived embedding potential rather than by using an isolated cluster. Second, the active space must be compact enough for near-term quantum resources while remaining chemically complete along the reaction coordinate; MBECAS-SR addresses this by selecting adsorbate and surface orbitals associated with bond rearrangement and adsorbate-metal coupling and by maintaining orbital consistency between initial, transition and final states. Third, the quantum-computing output must be converted into quantitatively reliable reaction or adsorption energies; in QC-DFET, hardware-sampled QSCI provides a variational determinant subspace, while SC-NEVPT2-based corrections recover residual dynamic correlation and reduce the sensitivity of final energies to imperfect quantum-computing wave functions.

The three applications across Cu(111) and Cu(100) test complementary aspects of this strategy. $H_2$ dissociation/desorption provides a stringent bidirectional barrier benchmark, where QC-DFET reaches near-chemical accuracy for both $H_2$ activation and associative desorption. CO adsorption tests the ability of the method to resolve a site-preference error that has challenged semi-local DFT for decades. QC-DFET recovers the experimentally observed top-site preference on Cu(111), while the 40-qubit Cu(100) calculation further tests the transferability of this result across surface facets. Across both surfaces, the corrected energetics are associated with localized C–Cu σ bonding rather than net charge transfer alone.

Extending QC-DFET beyond the ideal metal terrace will be equally important. Real catalytic interfaces often involve low-coordinated sites, oxide-derived structures, metal-oxide boundaries, explicit solvent, interfacial electric fields and proton-coupled electron-transfer steps. These effects can alter both the active orbitals and the reaction thermodynamics. The DFET foundation of QC-DFET is well suited to such extensions because it separates the local many-body problem from the larger environment, but practical applications will require more robust treatments of solvation, electrode potential and finite-temperature surface dynamics.

Several limitations define clear directions for future work. The embedding potential is optimized at the DFT level and kept fixed during the correlated calculations. A self-consistent correlated embedding cycle, in which the quantum-computing density updates the embedding potential, could improve the description of charge redistribution between the active cluster and the environment. The current MBECAS-SR protocol is chemically guided and should be further automated using orbital entanglement, response descriptors or energy-based selection criteria, especially for reaction networks containing many intermediates and transition states. Dynamic correlation is presently introduced by a classical SC-NEVPT2-based correction; incorporating dynamic correlation more directly into quantum-computing workflows, or developing embedding-compatible quantum/classical perturbative schemes, would make the methodology more internally consistent.

Overall, the present results show that quantum computing can already play a chemically meaningful role in surface-reaction modelling when embedded in an appropriate many-body framework. By combining environment-aware DFET Hamiltonians, reaction-consistent active spaces, hardware-sampled QSCI and perturbative dynamic-correlation recovery, QC-DFET links near-term quantum computation to experimentally constrained catalytic energetics. This establishes a practical route for using quantum hardware not as an isolated solver for small molecular models, but as a targeted many-body component within predictive simulations of realistic surface chemistry.

## Methods

This section describes the QC-DFET implementation used to construct and solve environment-embedded active-space Hamiltonians for surface reactions. The workflow consists of four components: DFET embedding-potential optimization, MBECAS-SR active-space construction, quantum-computing solution of the embedded active-space Hamiltonian, and perturbative dynamic-correlation correction for final energy evaluation. Detailed computational settings, system-specific active-space selections, and quantum-hardware and circuit details are provided in the Supplementary Information.

**Density-functional embedding theory**

The embedding potential is optimized using the Wu-Yang density-matching formulation.[51] The full DFT density is partitioned into cluster and environment contributions, and the embedding potential is determined by maximizing the Wu-Yang functional $W[V_{\mathrm{emb}}]$

$$W[V_{\mathrm{emb}}] = E_{\mathrm{cl}}[\rho_{\mathrm{cl}}, V_{\mathrm{emb}}] + E_{\mathrm{env}}[\rho_{\mathrm{env}}, V_{\mathrm{emb}}] - \int \rho_{\mathrm{tot}} V_{\mathrm{emb}}\, d\boldsymbol{r} \tag{1}$$

where $E_{\mathrm{cl}}$ and $E_{\mathrm{env}}$ denote the DFT energies of the cluster and environment in the presence of the embedding potential $V_{\mathrm{emb}}$, $\rho_{\mathrm{cl}}$ and $\rho_{\mathrm{env}}$ are the corresponding subsystem densities, and $\rho_{\mathrm{tot}}$ is the target density obtained from the DFT calculation of the full surface system. The stationary condition with respect to the embedding potential gives

$$\rho_{\mathrm{tot}} = \rho_{\mathrm{cl}} + \rho_{\mathrm{env}} \tag{2}$$

which enforces density matching between the full system and the embedded subsystems. For practical embedded cluster calculations, the optimized embedding potential is represented in the Gaussian-type-orbital (GTO) basis. Its matrix elements are evaluated as

$$V_{pq}^{\mathrm{emb}} = \langle \chi_p | \hat{V}_{\mathrm{emb}} | \chi_q \rangle \tag{3}$$

where $\chi_p$ and $\chi_q$ are GTO basis functions. This embedding contribution is added to the one-electron part of the cluster Hamiltonian, yielding the embedded Hamiltonian

$$\hat{H}_{\mathrm{emb}} = \sum_{p,q} \left(h_{pq} + V_{pq}^{\mathrm{emb}}\right) a_p^{\dagger} a_q + \frac{1}{2} \sum_{p,q,r,s} \langle pq|rs\rangle a_p^{\dagger} a_q^{\dagger} a_s a_r \tag{4}$$

The resulting Hamiltonian retains the cluster–environment interactions through $V_{\mathrm{emb}}$ while keeping explicit two-electron interactions within the embedded cluster. It defines the embedded DFT reference and, after active-space projection, the many-body Hamiltonian solved by the quantum-computing solvers. In the present work, following previous DFET studies of surface reactions, the embedding potential is optimized at the DFT level and kept fixed during the correlated and quantum-computing calculations.[34,35] This choice isolates the solver-level many-body correction while retaining the cluster-environment interaction encoded in the DFT-derived embedding potential.

**Reaction-consistent active-space construction**

The embedded Hamiltonian is reduced to a compact active-space problem using MBECAS-SR, a chemically guided selection protocol for surface reactions. Chemically important molecular-orbital pairs are first chosen as initial core orbitals according to bonding patterns along the reaction coordinate, including bond breaking, bond formation and adsorbate-surface coupling. Starting from these core orbitals, MBECAS-SR identifies important substrate orbitals and missing adsorbate orbitals, with the final active space controlled by a single tunable screening parameter. Orbital consistency along the potential-energy surface is maintained through a creeping strategy, in which the converged orbitals from one structure are propagated as the initial guess for the subsequent structure. Further implementation details and system-specific active-space selections are provided in Supplementary **Section 2**.

After active-space construction, the embedded Hamiltonian is projected onto the selected active-orbital manifold, yielding the embedded active-space Hamiltonian

$$\hat{H}_{\mathrm{act}}^{\mathrm{emb}} = \sum_{p,q \in \mathrm{act}} h_{pq}^{\mathrm{emb,act}}\, a_p^{\dagger} a_q + \frac{1}{2} \sum_{p,q,r,s \in \mathrm{act}} g_{pqrs}^{\mathrm{act}}\, a_p^{\dagger} a_q^{\dagger} a_s a_r \tag{5}$$

where $h_{pq}^{\text{emb,act}}$ and $g_{pqrs}^{\text{act}}$ denote the one- and two-electron integrals of $\hat{H}_{\text{emb}}$ projected into the selected active orbital space with the embedding contribution retained in the one-electron term through $V_{pq}^{\text{emb}}$. This active-space Hamiltonian defines the many-body problem solved by the quantum-computing solvers.

**Quantum-computing active-space solvers**

We use hardware-sampled QSCI experiments[47,48] to solve the embedded active-space Hamiltonians. Low-depth circuits are constructed from early-stage operators selected by ADAPT-UCCGSD[40,43]. The single- and double-excitation operators selected up to ADAPT steps 5 and 10 are separately mapped to fermionic-excitation circuits and used to prepare correlated trial states on quantum hardware. The excitation circuits follow compact fermionic-excitation constructions without independently expanding each excitation into Pauli-string evolution circuits.[44] Measurements of the hardware-prepared trial states in the computational basis provide determinant samples for the subsequent QSCI procedure. Detailed circuit-construction procedures and circuit-resource analyses are provided in Supplementary **Section 5**.

QSCI is implemented as a hybrid quantum–classical selected-CI procedure.[47,48] A correlated trial state $|\Psi(\boldsymbol{\theta})\rangle$ prepared on quantum hardware is measured in the computational basis, yielding bitstrings $\{\boldsymbol{b}\}$ with empirical probabilities $p(\boldsymbol{b})$. Each bitstring is mapped to a Slater determinant $|D_{\boldsymbol{b}}\rangle$, and a selected determinant subspace is constructed by retaining either the top K most frequently sampled determinants or all determinants above a probability threshold while enforcing the target alpha- and beta-electron numbers:

$$\mathcal{S} = \{|D_i\rangle : p_i \geq p_{\text{cut}}\} \quad \text{or} \quad \mathcal{S} = \text{TopK}(N_{\text{det}}) \tag{6}$$

Given the selected determinant set $\mathcal{S}$, the Hamiltonian matrix is formed in the truncated determinant subspace,

$$H_{ij} = \langle D_i|\hat{H}|D_j\rangle, \qquad |D_i\rangle, |D_j\rangle \in \mathcal{S} \tag{7}$$

and the QSCI energy is obtained by classical subspace diagonalization:

$$E_{\text{QSCI}} = \min_{\boldsymbol{c}} \frac{\boldsymbol{c}^\dagger \boldsymbol{H} \boldsymbol{c}}{\boldsymbol{c}^\dagger \boldsymbol{c}} \tag{8}$$

As the threshold is lowered, or K is increased, the selected CI space is systematically enlarged and the variational energy approaches the full-CI limit within the chosen active space. To improve robustness against hardware noise, a configuration-recovery procedure is further applied to restore physically valid electronic configurations from noisy measurement samples.[48] For the 40-qubit calculations, we introduce correlation-directed self-consistent configuration recovery (CD-SCORE), which incorporates one-body and conditional pair-occupation information into bitstring recovery and particle-number-conserving proposals. We further complement CD-SCORE with our previously developed H-couple[52] method for Hamiltonian-guided expansion of the variational determinant subspace (Supplementary Section 5.3).

In some test calculations, we also performed noiseless classical simulations of ADAPT-VQE quantum algorithms. Two fermionic excitation pools were considered: UCCSD[42] and UCCGSD[43].

**Dynamic-correlation correction and total-energy evaluation**

Dynamic correlation outside the active space is included perturbatively using strongly contracted N-electron valence state second-order perturbation theory (SC-NEVPT2)[53] for the 24-qubit calculations and its uncontracted Epstein–Nesbet perturbation theory approximation (EN-SC-NEVPT2)[54] for the 40-qubit calculations. For the latter, spin-pure selected-CI reference wave functions are constructed from QSCI-selected spatial configurations through spin adaptation and Hamiltonian rediagonalization (Supplementary Section 5.4). The final embedded quantum energy is evaluated as:

$$E^{\text{QC-DFET}} = E_{\text{tot}}^{\text{PW-DFT}} + (E_{\text{cl}}^{\text{emb-QC-NEVPT2}} - E_{\text{cl}}^{\text{emb-DFT}}) \tag{9}$$

Here, emb-QC and emb-DFT denote embedded quantum-computing and embedded DFT cluster calculations in the presence of the embedding potential. $E_{\text{tot}}^{\text{PW-DFT}}$ is the plane-wave DFT total energy of the periodic slab, with the D3 dispersion correction removed. $E_{\text{cl}}^{\text{emb-DFT}}$ was obtained from DFT calculations using the PBE exchange–correlation functional in the GTO basis. It should be noticed that the vibrational zero-point energy (ZPE) contribution, intrinsic to low-temperature surface reaction energetics, should be considered; therefore, ZPE corrections are included for the calculated electronic energies of all systems. For adsorption processes, an additional thermodynamic correction arising from the change in the number of gas-phase molecules, namely the $\Delta nRT$ term, is also taken into account when converting electronic energies to adsorption enthalpies.

## Acknowledgements

The authors thank Prof. Emily A. Carter, Prof. Junwei Lucas Bao, and Prof. Ting Tan, and are particularly grateful to Prof. Qing Zhao and Dr. Wei Liu for their helpful discussions, suggestions and valuable guidance on DFET. This work is supported by the National Natural Science Foundation of China (22393913, 22303090, 22422304, 22073086), the Strategic Priority Research Program (XDB0450101) and the robotic AI-Scientist platform of the Chinese Academy of Sciences, and the USTC Supercomputing Center.

## References:

1. Nørskov, J. K., Abild-Pedersen, F., Studt, F. & Bligaard, T. Density functional theory in surface chemistry and catalysis. Proceedings of the National Academy of Sciences 108, 937–943 (2011).
2. Seh, Z. W. et al. Combining theory and experiment in electrocatalysis: insights into materials design. Science 355, eaad4998 (2017).
3. Birdja, Y. Y. et al. Advances and challenges in understanding the electrocatalytic conversion of carbon dioxide to fuels. Nature Energy 4, 732–745 (2019).
4. Kroes, G.-J. & Meyer, J. Best-of-both-worlds computational approaches to difficult-to-model dissociation reactions on metal surfaces. Chemical Science 16, 480–506 (2025).
5. Cohen, A. J., Mori-Sánchez, P. & Yang, W. Challenges for density functional theory. Chemical Reviews 112, 289–320 (2012).
6. Christensen, R., Hansen, H. A. & Vegge, T. Identifying systematic DFT errors in catalytic reactions. Catalysis Science & Technology 5, 4946–4949 (2015).
7. Verma, P. & Truhlar, D. G. Status and challenges of density functional theory. Trends in Chemistry 2, 302–318 (2020).
8. Cao, K., Füchsel, G., Kleyn, A. W. & Juurlink, L. B. F. Hydrogen adsorption and desorption from Cu (111) and Cu (211). Physical Chemistry Chemical Physics 20, 22477–22488 (2018).
9. Díaz, C. et al. Chemically accurate simulation of a prototypical surface reaction: $H_2$ dissociation on Cu (111). Science 326, 832–834 (2009).

10. Feibelman, P. J. et al. The CO/Pt (111) puzzle. The Journal of Physical Chemistry B 105, 4018–4025 (2001).
11. Patra, A., Peng, H., Sun, J. & Perdew, J. P. Rethinking CO adsorption on transition-metal surfaces: effect of density-driven self-interaction errors. Physical Review B 100, 035442 (2019).
12. Wellendorff, J., Silbaugh, T. L., Garcia-Pintos, D., Nørskov, J. K., Bligaard, T., Studt, F. & Campbell, C. T. A benchmark database for adsorption bond energies to transition metal surfaces and comparison to selected DFT functionals. Surface Science 640, 36–44 (2015).
13. Vollmer, S., Witte, G. & Wöll, C. Determination of site specific adsorption energies of CO on copper. Catalysis Letters 77, 97–101 (2001).
14. Wei, Z., Göltl, F. & Sautet, P. Diffusion barriers for carbon monoxide on the Cu(001) surface using many-body perturbation theory and various density functionals. Journal of Chemical Theory and Computation 17, 7862–7872 (2021).
15. Zhao, Q. & Carter, E. A. Revisiting competing paths in electrochemical $CO_2$ reduction on copper via embedded correlated wavefunction theory. Journal of Chemical Theory and Computation 16, 6528–6538 (2020).
16. Hsing, C.-R., Chang, C.-M., Cheng, C. & Wei, C.-M. Quantum Monte Carlo studies of CO adsorption on transition metal surfaces. The Journal of Physical Chemistry C 123, 15659–15664 (2019).
17. Chen, Z., Liu, Z. & Xu, X. Accurate descriptions of molecule-surface interactions in electrocatalytic $CO_2$ reduction on the copper surfaces. Nature Communications 14, 936 (2023).
18. Blyholder, G. Molecular orbital view of chemisorbed carbon monoxide. The Journal of Physical Chemistry 68, 2772–2777 (1964).
19. Shi, B. X. et al. Many-body methods for surface chemistry come of age: achieving consensus with experiments. Journal of the American Chemical Society 145, 25372–25381 (2023).
20. Zhang, R., Hao, X., Duan, T. & Wang, B. Adsorption and activation of CO and $H_2$, the corresponding equilibrium phase diagrams under different temperature and partial pressures over Cu(100) surface: Insights into the effects of coverage and solvent effect. Fuel Processing Technology 156, 253–264 (2017).
21. Begin, E. & Bao, J. L. Multiconfiguration Pair-Density Functional Theory with Quantum Embedding Predicts Correct CO Adsorption Sites on Copper Facets. The Journal of Physical Chemistry Letters 16, 9977–9984 (2025).
22. Huang, Z. et al. Advancing surface chemistry with large-scale ab-initio quantum many-body simulations. arXiv preprint arXiv:2412.18553 (2024).
23. Cao, C. et al. Quantum many-body simulations of catalytic metal surfaces. arXiv preprint arXiv:2508.13036 (2025).
24. Bartlett, R. J. & Musiał, M. Coupled-cluster theory in quantum chemistry. Reviews of Modern Physics 79, 291–352 (2007).
25. Chan, G. K.-L. & Sharma, S. The density matrix renormalization group in quantum chemistry. Annual Review of Physical Chemistry 62, 465–481 (2011).
26. Levine, D. S. et al. CASSCF with extremely large active spaces using the adaptive sampling configuration interaction method. Journal of Chemical Theory and Computation 16, 2340–2354 (2020).
27. Jones, L. O., Mosquera, M. A., Schatz, G. C. & Ratner, M. A. Embedding methods for quantum chemistry: applications from materials to life sciences. Journal of the American Chemical Society 142, 3281–3295 (2020).
28. Li, W. et al. Toward practical quantum embedding simulation of realistic chemical systems on near-term quantum computers. Chemical Science 13, 8953–8962 (2022).
29. Cao, C. et al. Ab initio quantum simulation of strongly correlated materials with quantum embedding. npj Computational Materials 9, 78 (2023).
30. Govind, N., Wang, Y. A. & Carter, E. A. Electronic-structure calculations by first-principles density-based embedding of explicitly correlated systems. The Journal of Chemical Physics 110, 7677–7688 (1999).
31. Klüner, T., Govind, N., Wang, Y. A. & Carter, E. A. Periodic density functional embedding theory for complete active space self-consistent field and configuration interaction calculations: ground and excited states. The Journal of Chemical Physics 116, 42–54 (2002).
32. Libisch, F., Huang, C. & Carter, E. A. Embedded correlated wavefunction schemes: theory and

applications. Accounts of Chemical Research 47, 2768–2775 (2014).
33. Yu, K., Libisch, F. & Carter, E. A. Implementation of density functional embedding theory within the projector-augmented-wave method and applications to semiconductor defect states. The Journal of Chemical Physics 143, 102806 (2015).
34. Zhao, Q., Zhang, X., Martirez, J. M. P. & Carter, E. A. Benchmarking an embedded adaptive sampling configuration interaction method for surface reactions: $H_2$ desorption from and $CH_4$ dissociation on Cu (111). Journal of Chemical Theory and Computation 16, 7078–7088 (2020).
35. Zhao, Q., Martirez, J. M. P. & Carter, E. A. Revisiting understanding of electrochemical $CO_2$ reduction on Cu (111): competing proton-coupled electron transfer reaction mechanisms revealed by embedded correlated wavefunction theory. Journal of the American Chemical Society 143, 6152–6164 (2021).
36. Cao, Y. et al. Quantum chemistry in the age of quantum computing. Chemical Reviews 119, 10856–10915 (2019).
37. Liu, J., Fan, Y., Li, Z. & Yang, J. Quantum algorithms for electronic structures: basis sets and boundary conditions. Chemical Society Reviews 51, 3263–3279 (2022).
38. Zeng, X., Fan, Y., Liu, J., Li, Z. & Yang, J. Quantum neural network inspired hardware adaptable ansatz for efficient quantum simulation of chemical systems. Journal of Chemical Theory and Computation 19, 8587–8597 (2023).
39. Zeng, X. et al. Accurate chemical reaction modeling on a noisy intermediate-scale quantum computer with an active space-based workflow. The Journal of Physical Chemistry Letters 16, 11709–11716 (2025).
40. Grimsley, H. R., Economou, S. E., Barnes, E. & Mayhall, N. J. An adaptive variational algorithm for exact molecular simulations on a quantum computer. Nature Communications 10, 3007 (2019).
41. Jordan, P. & Wigner, E. Über das Paulische Äquivalenzverbot. Zeitschrift für Physik 47, 631–651 (1928).
42. Romero, J. et al. Strategies for quantum computing molecular energies using the unitary coupled cluster ansatz. Quantum Science and Technology 4, 014008 (2019).
43. Lee, J., Huggins, W. J., Head-Gordon, M. & Whaley, K. B. Generalized unitary coupled cluster wave functions for quantum computation. Journal of Chemical Theory and Computation 15, 311–324 (2018).
44. Yordanov, Y. S., Arvidsson-Shukur, D. R. M. & Barnes, C. H. W. Efficient quantum circuits for quantum computational chemistry. Physical Review A 102, 062612 (2020).
45. Zheng, Y. et al. Perturbative Variational Quantum Eigensolver via Reduced Density Matrices. The Journal of Physical Chemistry A 130, 8482–8492 (2026).
46. Zeng, X. et al. Transformer refined quantum sampling for strongly correlated electronic structure. arXiv preprint arXiv:2605.24617 (2026).
47. Kanno, K. et al. Quantum-selected configuration interaction: classical diagonalization of Hamiltonians in subspaces selected by quantum computers. Physical Review Research 8, 023268 (2026).
48. Robledo-Moreno, J. et al. Chemistry beyond the scale of exact diagonalization on a quantum-centric supercomputer. Science Advances 11, eadu9991 (2025).
49. Gujarati, T. P. et al. Quantum computation of reactions on surfaces using local embedding. npj Quantum Information 9, 88 (2023).
50. Barroca, M. A., Neumann Barros Ferreira, R. & Steiner, M. Exploring qubit-ADAPT-VQE for materials discovery in direct air capture. APL Quantum 1, 046103 (2024).
51. Wu, Q. & Yang, W. A direct optimization method for calculating density functionals and exchange–correlation potentials from electron densities. The Journal of Chemical Physics 118, 2498–2509 (2003).
52. Zhang, H., Zeng, X., Li, Z. & Zhou, Y. Hamiltonian-Guided Autoregressive Selected-Configuration Interaction Achieves Chemical Accuracy in Strongly Correlated Systems. Journal of Chemical Theory and Computation 21, 12622–12633 (2025).
53. Angeli, C., Cimiraglia, R. & Malrieu, J.-P. N-electron valence state perturbation theory: a fast implementation of the strongly contracted variant. Chemical Physics Letters 350, 297–305 (2001).
54. Ugandi, M. & Roemelt, M. An Uncontracted Epstein–Nesbet Perturbation Theory Approximation to SC-NEVPT2 Based on Spin-Pure Selected CI Wave Functions. Journal of Chemical Theory and Computation 22, 7164–7176 (2026).

# Supplementary Information for Embedded quantum computing for many-body surface reaction

Dedong Wan[1,2], Xiaopeng Li[1], Yi Fan[1], Jie Liu[3], Xiongzhi Zeng[1]*, Zhenyu Li[1]*

[1]State Key Laboratory of Precision and Intelligent Chemistry, University of Science and Technology of China, Hefei 230026, China

[2]Hefei National Research Center for Physical Sciences at the Microscale, University of Science and Technology of China, Hefei, 230026, China

[3] Hefei National Laboratory, University of Science and Technology of China, Hefei 230088, China

*To whom correspondence should be addressed:

xzzeng@ustc.edu.cn; zyli@ustc.edu.cn;

## 1. Computational details

### 1.1 Periodic DFT Calculations

Periodic Kohn–Sham DFT calculations were performed using the Vienna *Ab initio* Simulation Package[1,2] (VASP, version 6.4.2) with the all-electron frozen-core projector augmented-wave (PAW)[3] method and the PBE exchange–correlation functional[4]. Grimme's D3 dispersion correction[5,6] with the Becke–Johnson damping[7] function was included throughout. The $1s$ electron of H, the $2s$ and $2p$ electrons of C and O, and the $4s$ and $3d$ electrons of Cu were treated self-consistently. For $H_2$ dissociation/desorption and CO adsorption, the Cu(111) surface was modeled by a $\sqrt{21} \times \sqrt{21}$ four-layer slab containing 84 Cu atoms and the Cu(100) surface was modeled by a $5 \times 5$ four-layer slab containing 100 Cu atoms. A vacuum region of at least 15 Å was included in the periodic cell. During structural optimization, the atoms in the top two layers were relaxed, while those in the bottom two layers were fixed at their bulk positions. Geometry optimizations were continued until the maximum residual force on each atom was smaller than 0.03 eV/Å. A plane-wave basis set with a kinetic-energy cutoff of 660 eV was employed. To sample the Brillouin zone, Γ-centered Monkhorst–Pack[8] $14 \times 14 \times 14$, $4 \times 4 \times 1$, and $3 \times 3 \times 1$ $k$-point grids were used for the bulk unit cell and the $\sqrt{21} \times \sqrt{21}$ and $5 \times 5$ slab supercells, respectively. Electronic occupations were treated using the Methfessel–Paxton method[9] with a smearing width of 0.09 eV to facilitate self-consistent-field convergence. The minimum energy paths (MEPs) were optimized using the climbing-image nudged elastic band (CI-NEB)[10] method with an artificial spring-force constant of 3 eV/$\text{Å}^2$ along the reaction path. Vibrational frequencies ($\nu_j$) were computed at the PBE-D3BJ level using the periodic slab model for the MEP structures in order to verify the correctness of the transition states and to evaluate the zero-point energy (ZPE) corrections.

### 1.2 DFET embedding potentials and embedded cluster calculations

For $H_2$ dissociation/desorption, we adopt the embedding potential reported by Zhao et al[11] to enable direct comparison with their results. For the main CO adsorption calculations, two Cu14 clusters were carved separately from 5 × 5 four-layer Cu(111) and Cu(100) slabs, each containing 100 Cu atoms, while a $Cu_{12}$ cluster carved from the same 5 × 5 four-layer Cu(111) slab model was used for the cluster-size robustness analysis of CO adsorption, as shown in **Fig. S1**. The corresponding embedding potential $\hat{V}_{\text{emb}}$ was obtained using a modified version of VASP 6.4.2[12] and subsequently transformed from the real-space grid representation to an atomic Gaussian-type-orbital (GTO) basis using the EmbeddingIntegralGenerator code[13].

Embedding potentials are optimized using the PBE exchange–correlation functional. PySCF 2.7.0[14,15] was then used to construct the embedded Hamiltonian and its active-space projection in the GTO basis. For the CAS(22e,20o) calculations, orbital optimization was performed using the heat-bath selected configuration interaction self-consistent-field (SHCISCF) method[16], with the Dice selected-CI solver[17] interfaced through PySCF. The SHCI variational selection threshold was set to $\varepsilon = 1 \times 10^{-6}$ Ha. Dynamic-correlation corrections were evaluated using strongly contracted N-electron valence state second-order perturbation theory (SC-NEVPT2) as implemented in PySCF for the 24-qubit calculations. For the 40-qubit calculations, EN-SC-NEVPT2 corrections were evaluated using the HUMboldt Multi-Reference (HUMMR) program[18,19], with spin-pure selected-CI reference wave functions derived from QSCI configurations. The HUMMR screening parameters 'EpsilonHCIENEVPT2' and 'EpsilonHCINEVPT2' were both set to $1 \times 10^{-7}$. Further details of the QSCI–EN-SC-NEVPT2 interface are provided in Supplementary Section 5.4. The molecular geometries used in the embedded cluster calculations were generated by combining the embedded Cu clusters with adsorbates fixed at their periodic PBE-D3BJ-optimized positions. For the basis sets, we used the all-electron aug-cc-pVDZ basis for C, H, and O[20], together with a cc-pVDZ valence basis for Cu[21] and effective core potentials (ECPs) [22] replacing the 10 core electrons.

The active space selected for $H_2$ dissociation/desorption on Cu (111) was CAS (12e,12o), including two H-derived $\sigma, \sigma^*$ orbital pairs associated with H–H bond formation/cleavage and H–Cu bonding/antibonding interactions, together with four occupied and four unoccupied Cu 4s-derived orbitals on the $\text{Cu}_{12}$ cluster **(Fig. S5)**. The active space selected for CO adsorption on Cu (111) was CAS (14e,12o), including occupied C–O $\sigma_{2s}$ and $\sigma_{2s}^*$ orbitals formed by C 2s and O 2s electrons, one C–O $\sigma, \sigma^*$ pair, two nearly pure C–O $\pi, \pi^*$ pairs, one additional C–O $\pi, \pi^*$ pair strongly hybridized with Cu 4s-derived surface character, and one occupied and one unoccupied Cu 4s-derived orbital on the $\text{Cu}_{14}$ cluster (**Fig. S11**). For cluster-size robustness analysis using the $Cu_{12}$ cluster, an additional pair of Cu 4s-derived orbitals is included to complete CO-Cu interaction to CAS (16e, 14o) (**Fig. S12**). Two active spaces, CAS(16e,14o) and CAS(22e,20o), were considered for CO adsorption on Cu(100). The former also included occupied C–O $\sigma_{2s}$ and $\sigma_{2s}^*$ orbitals formed by C 2s and O 2s electrons, one C–O $\sigma, \sigma^*$ pair, two nearly pure C–O $\pi, \pi^*$ pairs, and one additional

C–O $\pi, \pi^*$ pair strongly hybridized with Cu 4s-derived surface character; the rest were all Cu 4s-derived orbitals (**Fig. S16**). For the latter, the adsorbate orbitals lacked the one additional C–O $\pi, \pi^*$ pair strongly hybridized with Cu 4s-derived surface character, and the rest were also all Cu 4s-derived orbitals (**Fig. S17**).

# 2. MBECAS-SR active-space construction

For certain chemical problems, quantitatively reliable results can be obtained by correlating only a limited number of orbitals and electrons directly involved in the reaction process. Therefore, active-space construction is essential and should be as automated and physics-based as possible. MBECAS was recently proposed by us for molecular systems, where the highest occupied molecular orbital (HOMO) and the lowest unoccupied molecular orbital (LUMO) are used as the minimal active space, followed by the evaluation of first- and second-order increments ranked by their absolute magnitudes[23]. Here, we adopt the basic idea of MBECAS and extend it to surface reactions, leading to the MBECAS-SR scheme (**Fig. S2**).

### 2.1 Implementation procedure of MBECAS-SR

**1. Definition and verification of the initial core orbitals.** First, a mean-field calculation, such as HF or DFT, is performed to identify the core bonding orbitals of the adsorbed molecule that are directly involved in bond formation or cleavage. These orbitals can usually be recognized from chemically intuitive bonding patterns. CASCI(2e,2o) calculations are then carried out by scanning the virtual orbitals to locate the corresponding antibonding partners. The selected bonding and antibonding orbitals define the initial active-orbital set, denoted by

$$\mathcal{A}_0 = \mathcal{A}_{\text{core-bonding}} \cup \mathcal{A}_{\text{core-antibonding}} \tag{1}$$

A CASSCF calculation with this initial active space is subsequently performed to verify that the chemically intended core orbitals are retained after orbital optimization.

**2. MBECAS-SR screening at the reference structure.** MBECAS-SR($\mathcal{A}_0$) analysis is performed for the structure in which the adsorbed molecule is fully dissociated, referred to here as the IS. We choose the IS as the reference because it exhibits the strongest wavefunction complexity along the reaction pathway and therefore requires the most complete orbital description (**Fig. S7a**).

Following the many-body-expanded correlation-energy formulation introduced in our previous MBECAS work, the CASCI energy associated with the initial active space is used as the reference energy,

$$\epsilon_{\text{ref}} = E_{\text{CASCI}}(\mathcal{A}_0) \tag{2}$$

Let $\Omega$ denote the expansion space containing the orbitals outside $\mathcal{A}_0$ that are considered during screening. For each orbital $i \in \Omega$ and each orbital pair $i, j \in \Omega$, additional CASCI calculations are performed with the augmented active spaces,

$$\epsilon_i = E_{\text{CASCI}}(\mathcal{A}_0 \cup \{i\}), \qquad \epsilon_{ij} = E_{\text{CASCI}}(\mathcal{A}_0 \cup \{i, j\}) \tag{3}$$

The first- and second-order correlation-energy increments are then defined as

$$\Delta\epsilon_i = \epsilon_i - \epsilon_{\text{ref}} \tag{4}$$

$$\Delta\epsilon_{ij} = (\epsilon_{ij} - \epsilon_{\text{ref}}) - \Delta\epsilon_i - \Delta\epsilon_j \tag{5}$$

The corresponding many-body expansion may be written formally as

$$E_{\text{FCI}} = \epsilon_{\text{ref}} + \sum_i \Delta\,\epsilon_i + \sum_{i<j} \Delta\,\epsilon_{ij} + \cdots \tag{6}$$

In the present MBECAS-SR implementation, the screening parameter $r_\epsilon$ is applied directly to the absolute first- and second-order correlation-energy increments. The two criteria can be expressed together as

$$i \in \mathcal{A}_{\text{raw}} \quad \Leftrightarrow \quad |\Delta\epsilon_i| > r_\epsilon \quad \text{or} \quad \exists\, j \neq i \text{ such that } |\Delta\epsilon_{ij}| > r_\epsilon \tag{7}$$

When the second condition is satisfied, both orbitals $i$ and $j$ are included in the raw candidate set. Here, $r_\epsilon$ is an absolute correlation-energy-increment threshold, and all $r_\epsilon$ values reported below are given in hartree (Ha). But for convenience, the threshold $r_\epsilon$ is reported without units hereafter. For moderate initial core-orbital sets, values of approximately $0.001$–$0.002$ are generally sufficient to obtain a compact, high-quality active space, although the appropriate threshold window remains system dependent.

Starting from the IS reference, this threshold is scanned from larger to smaller values to determine the order in which orbitals enter the candidate space. This procedure identifies both important substrate orbitals and missing adsorbate orbitals. Newly selected occupied orbitals are retained, whereas the treatment of additional virtual orbitals follows a conditional pairing branch rather than a global occupied-virtual equality constraint:

**(i).** When an added occupied orbital has an admissible virtual partner, the first corresponding virtual orbital encountered during the threshold scan is retained together with the occupied orbital to form a bonding-antibonding pair. The occupied and virtual additions are therefore balanced for this paired branch.

**(ii).** If no suitable virtual partner appears within the scanned threshold range, the occupied orbital is retained alone, and no additional virtual orbital is introduced for that occupied orbital.

Thus, virtual orbitals are not added merely to enforce equality between the total numbers of occupied and virtual orbitals. A final CASSCF calculation is then performed to verify that the selected orbitals remain properly represented after orbital optimization.

**3. Propagation along the potential-energy surface.** To ensure robust and reaction-consistent convergence along the potential-energy surface (PES), a "creeping" strategy is employed. The converged CASSCF orbitals from the preceding geometry are used as the initial guess for the next geometry. The propagation begins from the fully dissociated IS, so that the most complete orbital manifold identified at the electronically most complex structure is carried consistently along the reaction coordinate (**Fig. S7a**).

It should be noted that, in the present MBECAS-SR scheme, orbital optimization and active-space solving are performed together within a CASSCF framework. The computational cost of orbital optimization itself is relatively modest compared with the active-space solver, as it is dominated by Fock-like operations with formal quartic scaling in system size, and density fitting or Cholesky-decomposition techniques can further reduce both the scaling and prefactors, making this step comparable in cost to mean-field procedures[24]. By contrast, the dominant cost arises from the active-space solver, which scales exponentially on classical computers. This bottleneck can in principle be alleviated by replacing the classical solver with quantum-computing approaches such as VQE and QSCI.

At the same time, our orbital-selection strategy does require a certain amount of chemically motivated prior knowledge to define the initial set of orbitals. We acknowledge that the present choice of initial orbitals still relies heavily on chemical intuition. Further development of the MBECAS-SR strategy will therefore be pursued in future work. In the present study, this protocol is used primarily as a practical guide for active-space construction.

## 2.2 Details of MBECAS-SR for different systems

For $H_2$ dissociation/desorption, MBECAS-SR starts from an H 1s-dominated bonding/antibonding core pair and yields a CAS (12e,12o) active space within a narrow screening-threshold window of 0.0001–0.0002. For the desorption segment, direct MBECAS-SR selection gives active orbitals that are consistent with those obtained by creeping with manually selected active orbitals[11] (**Fig. S3**), indicating that this relatively simple part of the reaction coordinate can be described without an additional creeping procedure. This consistency reflects the smooth evolution of the $H_2$-derived bonding/antibonding pair and the gradually changing H–Cu coupling during associative desorption. In contrast, the dissociation segment is more demanding because the active space must retain the surface-coupled orbital channels associated with two separated H atoms while following the cleavage of the H–H bond. Therefore, for the production QC-DFET calculations, the active orbitals were propagated from the fully dissociated structure to maintain these dissociated-H-centered components along the bond-breaking coordinate.

For CO adsorption on Cu(111), the initial orbitals comprise three pairs of CO π- and σ-bonding orbitals together with their antibonding partners. Starting from this compact set, MBECAS-SR automatically recovers two additional adsorbate σ-type orbitals centered on C and O, followed by the correlated Cu slab orbital pairs needed for a balanced molecule–surface description, leading to a CAS(14e,12o) active space over the threshold range 0.0008–0.0013. Notably, the same active-orbital manifold is obtained whether the top, fcc, hcp and abv structures are treated independently or generated by creeping from the top-site geometry, indicating a reaction-consistent active-space description for this relatively simple surface rearrangement. For the cluster-size robustness analysis using the $Cu_{12}$ cluster, MBECAS-SR gives a CAS(16e,14o) active space within a similar threshold range of 0.0006–0.0013.

For CO adsorption on Cu(100), MBECAS-SR yields a CAS(16e,14o) active space over the threshold range 0.0005–0.0013. Lowering the threshold to 0.00030–0.00035 incorporates additional correlated Cu–CO orbital pairs and expands the active space to CAS(22e,20o). In addition, MBECAS-SR can be accelerated using density fitting[25] together with automatically generated auxiliary basis sets[26], yielding essentially identical orbital-selection results while substantially reducing the computational cost.

### 2.3 Active-space-size dependence for CO adsorption on Cu(100)

The dependence of the CO/Cu(100) energetics on the correlated active space was further examined by comparing CAS(16e,14o) and CAS(22e,20o). The smaller CAS(16e,14o) space already gives a top-site adsorption energy of −0.56 eV, within the experimental range, while retaining the top site as the most stable configuration (**Table S2**). However, the bridge–top and hollow–top separations are relatively large, 0.61 and 0.98 eV, respectively. The corresponding bridge and hollow active spaces contain orbitals with pronounced boundary-Cu 3d character (**Fig. S16**), indicating enhanced sensitivity to the restricted correlated orbital manifold. Expanding the active space to CAS(22e,20o) incorporates additional surface-correlated orbitals (**Fig. S17**) and reduces the bridge–top and hollow–top separations to 0.367 and 0.752 eV, respectively. Together with the Cu(111) cluster-size analysis, these results show that edge-localized orbital artifacts can arise from either spatial truncation of the embedded cluster or truncation of the correlated orbital space.

## 3. Delocalization-error analysis

To characterize whether the electronic-response differences between emb-DFT and emb-QC are accompanied by the characteristic delocalization tendency of semilocal DFT, we performed two complementary analyses at the $H_2$/Cu(111) transition-state (TS) geometry. The first examines the complete real-space density rearrangement relative to a common neutral-atom promolecule, avoiding an atom-partition-dependent description. The second evaluates the curvature of the fractional-electron energy for the fixed embedded PBE Hamiltonian, providing an energetic diagnostic of fractional-charge delocalization error.

### 3.1 Real-space density rearrangement relative to a common neutral-atom promolecule

For each electronic-structure level M, with M = emb-DFT or emb-QC, the same promolecular reference density was constructed by superposing the spherically averaged densities of neutral atoms at the atomic positions of the TS geometry,

$$\rho_{promol}(r) = \Sigma_A {\rho_A}^0(r) \quad (8)$$

The density rearrangement for method M was then defined relative to this common reference as

$$\Delta {\rho_M}^{\mathrm{def}}(r) = \rho_M(r) - \rho_{promol}(r), \quad M \in \{emb-DFT, emb-QC\} \quad (9)$$

Positive values of the deformation density denote electron-density accumulation relative to the neutral-atom promolecule, whereas negative values denote depletion. Because the identical promolecular density is used for emb-DFT and emb-QC, the two maps can be compared directly on a common symmetric color scale. Importantly, this quantity is not a direct emb-QC-minus-emb-DFT density difference and should not be interpreted as a direct measure of the net slab-to-adsorbate charge transfer. Instead, it shows how each method reorganizes the electronic density from the same neutral-atom reference upon formation of the adsorbate-surface electronic structure.

### 3.2 Fractional-electron energetic delocalization-error analysis

A complementary energetic diagnostic was obtained from fractional-electron calculations. For the exact density functional, the total energy is piecewise linear between adjacent integer electron numbers. We therefore evaluated the embedded PBE energy E(N ± α) for fractional electron addition and removal, where N is the neutral electron number and $0 \le \alpha \le 1$. The TS nuclear geometry and embedding potential were kept fixed throughout these calculations, so that the curvature reflects the fractional-charge response of the same embedded PBE Hamiltonian.

The calculated fractional-electron energy change relative to the neutral state is

$$\Delta E_{\pm}(\alpha) = E(N \pm \alpha) - E(N) \quad (10)$$

The corresponding piecewise-linear reference between the adjacent integer-electron endpoints is

$$E_{linear}(N \pm \alpha) = (1 - \alpha)E(N) + \alpha E(N \pm 1) \quad (11)$$

or, equivalently, for the energy change relative to E(N),

$$\Delta {E_{\pm}}^{linear}(\alpha) = \alpha[E(N \pm 1) - E(N)] \quad (12)$$

We define the deviation from piecewise linearity as

$$D_{\pm}(\alpha) = E_{linear}(N \pm \alpha) - E(N \pm \alpha) = (1 - \alpha)E(N) + \alpha E(N \pm 1) - E(N \pm \alpha) \quad (13)$$

For an exactly piecewise-linear energy, the deviation defined above is zero for all α. With the present sign convention, a positive deviation means that the calculated fractional-electron state lies below the straight line connecting the adjacent integer-electron states. The fractional-charge state is therefore artificially stabilized, which is the conventional energetic signature of delocalization error. A larger positive deviation corresponds to a stronger energetic preference for fractional charge.

# 4. Wavefunction analysis

### 4.1 Orbital-resolved Mayer bond-order analysis for CO–Cu bonding

To analyze how the correlated embedded quantum treatment modifies the CO–Cu bonding pattern, we computed orbital-resolved Mayer-type bond-order descriptors from the one-particle density matrices obtained at the emb-DFT and emb-QC levels.

For each electronic-structure level $X$, where $X = \text{emb-DFT}$ or $X = \text{emb-QC}$, we used the spin-summed AO density matrix $D^X$ and the AO overlap matrix $S$. The emb-QC density matrix was transformed back to the AO representation for the Mayer analysis. Since the selected active space contains the CO-centered and CO–Cu hybridized bonding and antibonding orbitals, including the $\pi/\pi^*$-type channels, the difference between the emb-QC and emb-DFT density matrices reflects the correlation-induced redistribution within these chemically relevant bonding channels.

The Mayer population matrix was defined as

$$P^X = D^X S \tag{14}$$

For two AO subspaces $A$ and $B$, represented by projector matrices $W_A$ and $W_B$, respectively, the generalized Mayer block index was evaluated as

$$M_{AB}^X = \mathrm{Tr}(W_A P^X W_B P^X) \tag{15}$$

When $W_A$ and $W_B$ are atom-centered AO projectors, Eq. (15) reduces to the conventional Mayer bond-order expression,

$$M_{AB}^X = \sum_{\mu \in A} \sum_{\nu \in B} (D^X S)_{\mu\nu} (D^X S)_{\nu\mu} \tag{16}$$

Here, we use the same expression with orbital-resolved AO subspace projectors to separate the $\sigma$- and $\pi$-type components of the CO–Cu bonding pattern. The correlation-induced correction is defined as

$$\Delta M_{AB} = M_{AB}^{\text{emb-QC}} - M_{AB}^{\text{emb-DFT}} \tag{17}$$

A negative value of $\Delta M_{AB}$ therefore indicates that the corresponding bonding descriptor is reduced at the emb-QC level relative to emb-DFT.

The local coordinate system for CO was defined by the C–O bond axis. The unit vector along this axis was constructed as

$$\boldsymbol{e}_{\mathrm{CO}} = \frac{\boldsymbol{R}_{\mathrm{O}} - \boldsymbol{R}_{\mathrm{C}}}{|\boldsymbol{R}_{\mathrm{O}} - \boldsymbol{R}_{\mathrm{C}}|} \tag{18}$$

For the C and O atoms, the $p$-type AO subspaces were decomposed into a component parallel to the C–O bond and a transverse component perpendicular to it. For each Cartesian $p$-type AO shell, the local $\sigma$-type projector was defined as

$$W_{p_z} = \boldsymbol{e}_{\mathrm{CO}} \boldsymbol{e}_{\mathrm{CO}}^T \tag{19}$$

and the local $\pi$-type projector was defined as

$$W_{p_{xy}} = I_3 - \boldsymbol{e}_{\mathrm{CO}} \boldsymbol{e}_{\mathrm{CO}}^T \tag{20}$$

These shell-level projectors were embedded into the full AO space to define the C- and O-centered $p_z$ and $p_{xy}$ projector matrices. In practice, all relevant $p$-type AO components on C and O were included, because the active-space orbitals involved in CO–Cu bonding can contain contributions from multiple $p$-type AO functions in the Gaussian basis.

The four orbital-resolved Mayer channels used in the analysis were

$$\begin{aligned} M_{\sigma(\mathrm{C-O})} &= M\left(W_{\mathrm{C},p_z}, W_{\mathrm{O},p_z}\right) \\ M_{\sigma(\mathrm{C-Cu})} &= M\left(W_{\mathrm{C},p_z}, W_{\mathrm{Cu}}\right), \\ M_{\pi(\mathrm{C-O})} &= M\left(W_{\mathrm{C},p_{xy}}, W_{\mathrm{O},p_{xy}}\right) \\ M_{\pi(\mathrm{C-Cu})} &= M\left(W_{\mathrm{C},p_{xy}}, W_{\mathrm{Cu},3d}\right) \end{aligned} \tag{21}$$

Here, $W_{\mathrm{Cu}}$ denotes the AO projector onto all basis functions centered on the Cu atoms in the cluster, whereas $W_{\mathrm{Cu},3d}$ denotes the projector onto the Cu $3d$-type AO subspace. Thus, the C–Cu $\sigma$ channel includes all Cu AO contributions, while the $\pi$ back-donation channel is resolved specifically through the Cu $3d$ subspace, consistent with the conventional Blyholder-type interpretation of CO adsorption.

The resulting Mayer-type quantities are dimensionless bonding descriptors. They are used to compare how selected CO–Cu bonding channels change from emb-DFT to emb-QC and should not be interpreted as a direct decomposition of the adsorption energy.

To further support the validity of our analysis, the resulting picture is also consistent with recent RPA–PBE Mayer bond-order analyses[27], in which many-body correlation was found to leave the internal C–O σ bond nearly unchanged, weaken the C–Cu σ interaction, strengthen the internal C–O π component, and substantially suppress the Cu–CO π-back-donation component (**Fig. S14**).

### 4.2 CI-based wavefunction complexity diagnostics

To quantify the complexity of the embedded correlated wave function, we analyze the CI expansion of the active-space state,

$$|\Psi\rangle = \sum_{I} c_I \, |\Phi_I\rangle \tag{22}$$

where $|\Phi_I\rangle$ denotes a many-electron configuration in the active space and $c_I$ is the corresponding CI coefficient. The normalized CI probability associated with each configuration is defined as

$$p_I = \frac{|c_I|^2}{\sum_J |c_J|^2} \tag{23}$$

Based on this probability distribution, we use two diagnostics: the top CI weight and the CI Shannon entropy.

The top CI weight is defined as the largest CI probability in the expansion,

$$w_{\mathrm{top}} = \max_I p_I \tag{24}$$

A large $w_{\mathrm{top}}$ indicates that the wave function can be described primarily by one dominant configuration, whereas a smaller value indicates that the wave function is more broadly distributed over multiple configurations.

The CI Shannon entropy is defined as

$$S_{\mathrm{CI}} = -\sum_{I} p_I \ln p_I \tag{25}$$

This quantity measures the information content of the CI probability distribution. A small $S_{\mathrm{CI}}$ corresponds to a compact wave function that can be represented with relatively little configurational information, while a larger $S_{\mathrm{CI}}$ indicates that the wave function is distributed over more configurations and therefore requires more information to describe. Together, $w_{\mathrm{top}}$ and $S_{\mathrm{CI}}$ provide complementary measures of the compactness and complexity of the embedded correlated wave function along the reaction coordinate.

### 4.3 Frontier-orbital descriptor analysis

To characterize the evolution of adsorbate and Cu-slab character in the embedded active space, we analyzed the atomic-orbital composition of the emb-QC active orbitals. The orbital-composition analysis was performed using the active orbitals from the emb-QC calculation; when natural orbitals were available, the emb-QC natural orbitals were used.

For each active orbital $i$, the Mulliken-type AO gross contribution was first evaluated from the AO coefficient vector $\boldsymbol{C}_i$ and the AO overlap matrix $\boldsymbol{S}$ as

$$g_{\mu i} = \left| C_{\mu i} (\boldsymbol{S}\boldsymbol{C}_i)_{\mu} \right| \tag{26}$$

where $\mu$ labels an atomic orbital. The absolute value was used to avoid cancellation between positive and negative AO gross populations. The normalized contribution of AO $\mu$ to active orbital $i$ is then

$$\tilde{g}_{\mu i} = \frac{g_{\mu i}}{\sum_{\nu} g_{\nu i}} \tag{27}$$

Based on this normalized AO population, the adsorbate and Cu-slab contributions to active orbital $i$ were defined as

$$A_i = \sum_{\mu \in \mathrm{ads}} \tilde{g}_{\mu i}, \qquad M_i = \sum_{\mu \in \mathrm{Cu}} \tilde{g}_{\mu i} \tag{28}$$

where $A_i$ measures the adsorbate character and $M_i$ measures the Cu-slab character of the active orbital. These two quantities are the basic descriptors used to monitor how the active orbitals evolve from adsorbate-centered to metal-coupled character along the reaction coordinate.

To focus on the chemically most relevant frontier-like active orbitals, each active orbital was assigned a frontier weight. In the present analysis, the weight favors orbitals with strong adsorbate character and significant local bonding character,

$$\omega_i = A_i L_i (\sigma_i + \alpha \pi_i) \tag{29}$$

where $L_i$ is the local adsorbate $s$-like contribution, $\sigma_i$ and $\pi_i$ are the corresponding adsorbate $\sigma$- and $\pi$-like scores, and $\alpha = 0.15$ was used to include a small contribution from $\pi$-like character. The weights were normalized over all active orbitals,

$$\bar{\omega}_i = \frac{\omega_i}{\sum_j \omega_j} \tag{30}$$

If all $\omega_i$ values were numerically negligible, equal weights were used instead.

The four active orbitals with the largest normalized frontier weights were selected as the frontier subset. For any orbital descriptor $X_i$, such as the adsorbate contribution $A_i$ or the Cu-slab contribution $M_i$, the plotted value was computed as the weighted average over this top-4 frontier subset,

$$\langle X \rangle_{\text{frontier}} = \frac{\sum_{i \in \text{top4}} \bar{\omega}_i X_i}{\sum_{i \in \text{top4}} \bar{\omega}_i} \tag{31}$$

The quantities shown in **Fig. S7b** are therefore $\langle A \rangle_{\text{frontier}}$ and $\langle M \rangle_{\text{frontier}}$, corresponding to the weighted adsorbate and Cu-slab contributions of the frontier active orbitals, respectively.

# 5. Quantum implementation and QSCI post-processing

## 5.1 Quantum computing settings

ADAPT-UCCSD and ADAPT-UCCGSD simulations and circuit optimization were executed in our in-house $Q^2$Chemistry software[28]. The reduction in the NEVPT2 total-energy error was evaluated using intermediate ADAPT-UCCGSD states sampled at different active-space QC energy errors relative to FCI along the convergence trajectory (**Fig. S10**).

The quantum-hardware experiments were performed on two generations of the *Zuchongzhi* superconducting quantum processor at the China Telecom "Tianyan" Quantum Computing Laboratory. The 24-qubit experiments for $H_2$ dissociation/desorption and CO adsorption on Cu(111) were carried out on Tianyan-176-II, corresponding to Zuchongzhi 2.0. The device provides 66 qubits and 110 tunable couplers, with average $T_1$ and $T_2$ (CPMG) times of 23.55 and 9.05 μs, respectively. The average readout, single-qubit gate, and two-qubit gate error rates were 2.32%, 0.24%, and 1.61%, respectively. The 40-qubit experiments for CO adsorption on Cu(100) were carried out on Tianyan-287, corresponding to Zuchongzhi 3.0. The processor provides 105 data qubits and 182 coupling qubits. According to the platform calibration record dated November 5, 2026, the average $T_1$ and $T_2$ (CPMG) times were 50.37 and 38.76 μs, respectively, while the average readout, single-qubit gate, and two-qubit gate error rates were 1.68%, 0.06%, and 0.48%, respectively.

For $H_2$ dissociation/desorption on Cu(111), we validated the quantum-experimental results obtained with configuration recovery using both the step-5 and step-10 circuits. Both results achieve chemical accuracy, while the step-10 circuit exhibits consistently better performance (**Fig. S9a**). Therefore, all subsequent quantum experiments were performed using the corresponding step-10 circuits, with the detailed results presented in **Figs. S9 and S18**. For the smaller active-space hardware calculations, configuration recovery was performed using the configuration-recovery module in Qiskit Addon SQD[29]. The recovery procedure was carried out for at most five iterations, with the convergence threshold for the QSCI energy set to ($10^{-6}$) Ha. For the 40-qubit calculations, the hardware-derived configuration space was expanded using the correlation-directed self-consistent configuration recovery (CD-SCORE) and Hamiltonian-guided H-couple[30] refinement procedures described in Supplementary Section 5.3.

## 5.2 Hardware circuit construction and optimization

**Fig. S8a** shows the convergence of the ADAPT-UCCGSD energy error and the corresponding circuit resources at representative ADAPT steps. The ADAPT-UCCGSD ansatz systematically lowers the energy error during the iterative operator selection and eventually reaches the chemical-accuracy regime. In particular, the error decreases below the chemical-accuracy threshold of 0.043 eV after sufficient ADAPT iterations, demonstrating that the selected UCCGSD operator pool is capable of preparing a highly accurate correlated state for the embedded active-space Hamiltonian.

However, the circuit cost increases rapidly as more ADAPT operators are included. Although the fully converged ADAPT circuit gives the lowest error, its circuit depth reaches the order of $10^4$–$10^5$, with approximately $4 \times 10^4$ single-qubit gates and $3.7 \times 10^4$ two-qubit gates in the final step. Such a circuit is far beyond the practical depth accessible in the present hardware experiments. Therefore, for the hardware demonstrations, we selected the early step-5 and step-10 ADAPT circuits as shallow representative circuits. These two circuits contain the leading ADAPT-selected excitation operators while maintaining experimentally manageable depths. After circuit optimization, for the IS of $H_2$ dissociation/desorption on Cu (111), the step-5 circuit has a depth of 43 with 92 single-qubit gates and 72 two-qubit gates, whereas the step-10 circuit has a depth of 162 with 188 single-qubit gates and 172 two-qubit gates (**Fig. S8a**). These shallow circuits were then used to prepare correlated trial states for the QSCI calculations.

In our implementation, we use the selected excitation operators and optimized amplitudes from the ADAPT-VQE output as the direct input for circuit construction. For a given ADAPT step, we read the corresponding list of selected operators and append the associated single- or double-fermionic-excitation circuit modules sequentially, following the order in which the operators are selected by ADAPT-VQE. The circuit construction follows the efficient fermionic-excitation strategy under the Jordan–Wigner mapping, where a compact qubit-excitation core is supplemented by a Jordan–Wigner parity correction to recover the required fermionic anticommutation sign. This avoids expanding each Pauli string independently with a full CNOT-staircase construction and thereby reduces the two-qubit-gate overhead.

For a single excitation, we use the input format [k, i], corresponding to the generator

$$a_k^\dagger a_i - a_i^\dagger a_k \quad (32)$$

with the ordering condition $i < k$. When the two spin orbitals are adjacent, we directly apply the corresponding two-qubit excitation module. When intermediate spin orbitals are present, we first encode the parity of the Jordan–Wigner string $\{i+1, \dots, k-1\}$ onto an anchor qubit using a CNOT staircase. We then apply controlled-phase operations to introduce the parity-dependent sign correction required by fermionic anticommutation, and finally uncompute the parity encoding by reversing the CNOT staircase. In this way, the single-excitation module retains the compact qubit-excitation structure while explicitly including the fermionic sign information from the Jordan–Wigner string.

For a double excitation, we use the input format [k, l, i, j], corresponding to the generator

$$a_k^\dagger a_l^\dagger a_i a_j - a_i^\dagger a_j^\dagger a_k a_l \quad (33)$$

The present circuit module is constructed for the canonical index ordering $i < j < k < l$. Since the hardware demonstrations in this work use the early UCCGSD-based ADAPT-VQE circuits, specifically the step-5 and step-10 circuits, the selected operators generally remain close to a simple occupied/virtual pair structure. We therefore use a restricted canonicalization procedure: index swaps are allowed only within the $(i, j)$ pair or within the $(k, l)$ pair. According to fermionic antisymmetry, each such internal pair swap changes the sign of the excitation generator; therefore, we multiply the excitation amplitude $\theta$ by $-1$ when an odd number of internal pair swaps is performed. If the excitation still does not satisfy $i < j < k < l$ after this pair-internal canonicalization, we do not apply a more general four-index reordering in the present simplified circuit module.

The double-excitation circuit is assembled in three steps. First, we use a short sequence of CNOT operations to establish the parity structure of the occupied and virtual orbital pairs. Second, we append the central controlled-rotation block that realizes the double-excitation core. Third, we encode the total parity of the two intermediate Jordan–Wigner segments,

$$\{i+1, \dots, j-1\} \quad \text{and} \quad \{k+1, \dots, l-1\} \quad (34)$$

and apply controlled-phase operations to impose the corresponding fermionic sign correction. After the parity correction, all CNOT staircases used for parity encoding are reversed so that no auxiliary parity information remains in the circuit.

After circuit construction, we perform gate-level optimization and resource analysis within the $Q^2$Chemistry workflow. The generated circuit is converted into a $Q^2$Chemistry-compatible circuit object, and the built-in circuit optimization routine is applied to simplify the gate sequence. We record the circuit depth, total number of gates, number of single-qubit gates, and number of two-qubit gates before and after optimization. The circuit resources reported for the step-5 and step-10 hardware experiments correspond to the optimized circuits obtained after this $Q^2$Chemistry optimization.

### 5.3 Configuration recovery and Hamiltonian-guided refinement for the 40-qubit calculations

Quantum measurements on present-day quantum processors are affected by hardware noise, which can redistribute measured probability toward configurations with little weight in the target quantum state. In particular, noise may produce bitstrings that violate the particle-number symmetries of the Hamiltonian, while finite sampling limits the coverage of relevant configurations. Configuration recovery addresses these effects using conserved quantities and occupation information obtained from variational Hamiltonian diagonalization. To improve the 40-qubit calculations, we propose correlation-directed self-consistent configuration recovery (CD-SCORE), which exploits one-body and conditional pair-occupation information, and complement it with our previously developed Hamiltonian-guided H-couple refinement to expand the variational determinant subspace toward convergence.

#### 5.3.1 Configuration-recovery methods

We denote the current determinant subspace by $P$, with the variational wavefunction

$$|\Psi_P\rangle = \sum_{I\in P} c_I \, |D_I\rangle. \tag{35}$$

Here $D_I$ is a Slater determinant and $c_I$ is its expansion coefficient, obtained by diagonalizing the projected Hamiltonian $H_{PP}$. The updated coefficients are used in subsequent configuration-recovery steps.

Sample-based quantum diagonalization (SQD)[31] is a quantum–classical workflow that combines configuration recovery, projected Hamiltonian diagonalization, and self-consistent updates of orbital occupations. Its self-consistent configuration recovery (S-CORE) procedure uses average orbital occupations from the current variational wavefunction to guide stochastic correction of measured bitstrings with incorrect spin-resolved particle numbers. Occupied bits are removed or unoccupied bits are added with occupation-biased probabilities until the required particle-number sector is restored; measured bitstrings already in the correct sector are left unchanged by this correction. The recovered configurations enter subspace diagonalization, providing updated occupation estimates for later recovery cycles.

The central idea of CD-SCORE is to incorporate correlations between orbital occupations into the recovery process, beyond the marginal occupation statistics used in S-CORE. Specifically, CD-SCORE combines one-body occupation preferences with conditional pair-occupation correlations to guide bitstring correction and particle-number-conserving proposals. These occupation statistics are evaluated using measured configurations with the required electron numbers, whereas the complete set of original measured bitstrings is retained for configuration recovery and proposal generation. The corresponding target statistics are obtained from the current variational wavefunction. The one-body bias is defined as

$$b_p = \mathrm{logit}\left(n_p^T\right) - \mathrm{logit}\left(n_p^R\right). \tag{36}$$

Here $R$ denotes the measured bitstring distribution within the target electron-number sector and $T$ denotes the distribution of the current variational wavefunction, with weights proportional to $|c_I|^2$. Let $x_p \in \{0,1\}$ denote the occupation of spin orbital $p$ in a configuration $x$, and $w_D(x)$ its normalized weight in distribution $D \in \{R, T\}$. The mean occupation is $n_p^D = \sum_x w_D(x) x_p$. The log-odds function is $\mathrm{logit}(z) = \ln[z/(1-z)]$, with occupation probabilities clipped numerically at the endpoints.

Conditional pair-occupation information enters through the log-odds response

$$\chi_{p\leftarrow j}^D = \mathrm{logit}\left(n_{p|j}^D\right) - \mathrm{logit}\left(n_p^D\right). \tag{37}$$

Here $n_{p|j}^D = P_D\left(x_p = 1 \mid x_j = 1\right)$ is the probability that the orbital of interest is occupied given an occupied partner orbital, evaluated from the pair-occupation moment $\langle x_p x_j \rangle_D$ and the corresponding one-body occupations. The difference between the target and measured responses is used to construct the configuration-dependent pair field

$$g_{p,0}(x) = \frac{\sum_{j\neq p} \Delta\chi_{p\leftarrow j} x_j}{\sqrt{N_{\mathrm{partner},p}(x)}}. \tag{38}$$

Here $\Delta\chi_{p\leftarrow j} = \chi_{p\leftarrow j}^T - \chi_{p\leftarrow j}^R$ is the target-minus-measured conditional response and $N_{\mathrm{partner},p}(x)$ counts eligible occupied partner orbitals (with a minimum of one), excluding $j = p$. To stabilize the pair contribution relative to the one-body field, we use the scaling factor

$$s = \min\left(1, \frac{\rho_{\max}}{\rho}\right), \qquad \rho_{\max} = 0.5. \tag{39}$$

Here $\rho = \mathrm{RMS}(g_0)/\mathrm{RMS}(b)$ is the ratio of the RMS pair field to the RMS one-body bias; both RMS values are evaluated over the raw proposal ensemble. With the resulting local occupation field $\phi_p(x) = b_p + s\, g_{p,0}(x)$, the directional substitution score is

$$S_{p\to q}(x) = \phi_q(x) - \phi_p(x). \tag{40}$$

For a substitution from an occupied spin orbital $p$ to an empty spin orbital $q$, stochastic proposals are drawn with probabilities biased by $\exp\left[S_{p\to q}(x)/\tau\right]$, where $\tau$ is the proposal temperature. Bitstrings outside the required electron-number sector are corrected through occupation additions or removals directed by the local fields. Valid-electron-number strings are explored through spin-resolved occupied-to-empty substitutions, including permitted paired substitutions that conserve the electron counts. Newly generated configurations are deduplicated and selected under a fixed determinant-addition budget.

Hamiltonian-guided H-couple refinement expands the current determinant subspace through Hamiltonian-connected single- and double-substitution channels. For an external determinant $D_a$ and the current variational wavefunction, its coherent Hamiltonian coupling is

$$r_a = \left\langle D_a \middle| \hat{H} \middle| \Psi_P \right\rangle = \sum_{I\in P} H_{aI}\, c_I. \tag{41}$$

The external determinants are ranked according to

$$S_H(a) = |r_a|^2 = \left|\sum_{I\in P} H_{aI}\, c_I\right|^2. \tag{42}$$

Contributions $H_{aI} c_I$ from different parent determinants are summed coherently before their squared magnitude is evaluated. For candidate generation, determinants with significant CI weights form a parent set. Couplings to the same external determinant are accumulated across parent batches, and the highest-ranked candidates are selected for variational expansion.

At each configuration-selection stage, accepted candidates are merged with the current determinant support and deduplicated. Denoting the newly accepted determinant set at iteration $k$ by $\mathcal{A}_k$, the enlarged subspace is

$$P_{k+1} = P_k \cup \mathcal{A}_k. \quad (43)$$

The Hamiltonian is re-diagonalized in $P_{k+1}$, and the updated coefficients and occupation statistics inform subsequent configuration-recovery or Hamiltonian-guided refinement steps. Repeated selection and variational diagonalization yield the final 40-qubit wavefunctions.

### 5.3.2 Recovery results and discussion

Figure **S18** summarizes the configuration-recovery benchmarks and the variational convergence of the 40-qubit calculations. Panel (a) compares six rounds of S-CORE and CD-SCORE, while panel (b) assesses recovery from two independently measured bridge-site quantum datasets. Panel (c) compares the single-round energy reductions achieved by S-CORE, CD-SCORE, and H-couple under the same determinant-addition budget. Panels (d) and (e) present the top-site convergence trajectory and final errors across the four adsorption sites, respectively, and panel (f) examines the slowly converging tail of the hollow-site determinant space.

CD-SCORE achieves a substantially greater variational-energy reduction than S-CORE at a matched determinant-addition budget (**Fig. S18a**). With six successive additions of 50,000 determinants, the mean error falls from 207.176 mHa to 35.972 $\pm$ 0.547 mHa using CD-SCORE, compared with 167.479 $\pm$ 4.808 mHa using S-CORE. The stronger reduction demonstrates the benefit of pair-conditioned occupation information for configuration recovery.

The measured quantum-data quality also influences the attainable variational accuracy (**Fig. S18b**). The higher-yield and lower-yield datasets each contain 1,000,000 shots, with valid measurement fractions of 3.407% and 1.534%, respectively. Their initial variational errors are 212.436 and 1040.067 mHa. Following the same CD-SCORE procedure and the same 300,000-determinant expansion budget, the errors decrease to 31.220 $\pm$ 0.208 and 156.568 $\pm$ 23.264 mHa, respectively. The higher-yield quantum data thus retain a substantial accuracy advantage after recovery.

The single-round comparison in **Fig. S18c** evaluates three configuration-selection strategies at the same determinant-addition budget. For 50,000 newly added determinants, S-CORE lowers the variational energy by 8.869 $\pm$ 3.273 mHa, CD-SCORE by 60.748 $\pm$ 4.299 mHa, and H-couple by 179.840 mHa. CD-SCORE yields a 6.85-fold larger variational-energy reduction than S-CORE in this comparison, while H-couple provides the largest reduction.

Continued configuration recovery produces a systematic decrease in the 40-qubit variational error (**Fig. S18d,e**). In the representative convergence trajectory, the error first falls below the 1 kcal mol$^{-1}$ threshold at approximately 1.40 million determinants and ultimately reaches 1.054 mHa with 2,919,985 determinants. The final errors are 1.054, 1.560, 0.810, and 2.694 mHa for top, bridge, abv, and hollow, respectively. The first three are within the chemical-accuracy threshold, whereas the hollow configuration retains a larger residual error.

The hollow calculation exhibits slowly diminishing variational-energy gains during late-stage configuration recovery (**Fig. S18f**). The cumulative screened Epstein–Nesbet second-order estimate for the remaining determinants in the same 40-qubit configuration space grows as progressively larger external sets are included, approaching the scale of the remaining 2.694-mHa variational error. This behavior indicates that the missing contributions are distributed over a broad tail of weakly coupled configurations.

## 5.4 Spin-adapted QSCI interface to EN-SC-NEVPT2 for the 40-qubit calculations

### 5.4.1 Spin-pure reference construction and perturbative interface

For CO/Cu(100), hardware measurements and the recovery procedures in Section 5.3 yield a QSCI wave function expressed in selected Slater determinants. Determinants with the same spatial occupation pattern are grouped into configurations (CFGs), providing the candidate space for spin-pure selected-CI references.

Fixed electron number and spin projection do not guarantee a total-spin eigenstate. Each selected CFG is completed with all compatible singlet (S = 0) configuration state functions (CSFs), including spin couplings absent from the original QSCI determinants. This is not a coefficient-preserving transformation. The spin-pure reference is

$$\Psi_{\text{ref}}^{S=0} = \sum_{I\in C_{QSCI}} \sum_{\mu} C_{I\mu}\, \Phi_{I\mu}^{S=0} \tag{44}$$

Spin completion expands the full QSCI-derived CFG pools to over 21 million singlet CSFs for top and 18 million for abv, imposing substantial memory demands on rediagonalization and perturbation theory. We therefore select whole CFGs using their summed squared QSCI determinant coefficients, weighted inversely by the square root of the associated singlet-CSF counts. Every selected CFG retains all compatible singlet couplings. Increasing CSF budgets produce nested spaces, while full-CFG rediagonalization provides the comparison baseline.

The embedded Hamiltonian is rediagonalized in each selected singlet-CSF basis. Orbital ordering, molecular integrals, the core Fockian, and the perturbative orbital frame are maintained consistently across the interface. The projected eigenvalue equation is

$$H_{C}^{S=0}\, c = E_{\text{ref}}^{\text{SCI}}\, c \tag{45}$$

The rediagonalized spin-pure selected-CI reference, rather than the original QSCI coefficient vector, is passed to perturbation theory. The change in variational energy induced by spin completion or subsequent CFG selection is part of reference construction and is distinct from the second-order dynamical-correlation correction.

Dynamical correlation is evaluated using the EN-SC-NEVPT2[19] method. Six classes of perturbers retain the strongly contracted Dyall treatment, whereas the two computationally demanding classes use uncontracted Epstein–Nesbet perturbation theory. The residual contributions associated with an incomplete selected-CI reference are retained.

The corrected embedded energy obtained from the spin-pure reference is

$$E_{\text{emb}}^{\text{corr}} = E_{\text{ref}} + E_{\text{EN-SC}}^{(2)} \tag{46}$$

Here, the embedded reference energy includes the appropriate core and embedding contributions, while the second-order term provides the EN-SC-NEVPT2 dynamic-correlation correction.

### 5.4.2 Numerical results

Table S3 compares variational errors of the original QSCI states, the complete spin-adapted spaces constructed from the QSCI-derived configurations (CFGs), and the selected singlet-CSF reference spaces used for EN-SC-NEVPT2. All variational errors are evaluated relative to the corresponding high-accuracy SHCI reference energies. The complete CFG spaces contain 21,533,207, 18,168,193, 1,346,807, and 2,515,433 singlet CSFs for top, abv, bridge, and hollow, respectively.

For the reported EN-SC-NEVPT2 calculations, each of the four adsorption configurations uses a spin-pure reference of approximately one million singlet CSFs selected by the whole-CFG procedure described in Section 5.4.1. The selected-CSF variational errors are 1.120, 0.528, 1.195, and 3.269 mHa for top, abv, bridge, and hollow, respectively (Table S3). Using a common CSF budget provides a uniform reference-size protocol but does not ensure equal truncation errors across the four configurations.

Using the same four-layer 5 × 5 Cu(100) PBE energies without dispersion and applying zero-point and gas-phase thermodynamic corrections, QC-DFET predicts a top-site CO adsorption energy of −0.588 eV, with bridge–top and hollow–top relative energies of +0.364 and +0.761 eV, respectively (Fig. 6). The top-site preference is preserved. These energies are results for specified finite reference spaces and are not extrapolations to the full-CFG or full-CI limits.

## 6. Supplementary Figures

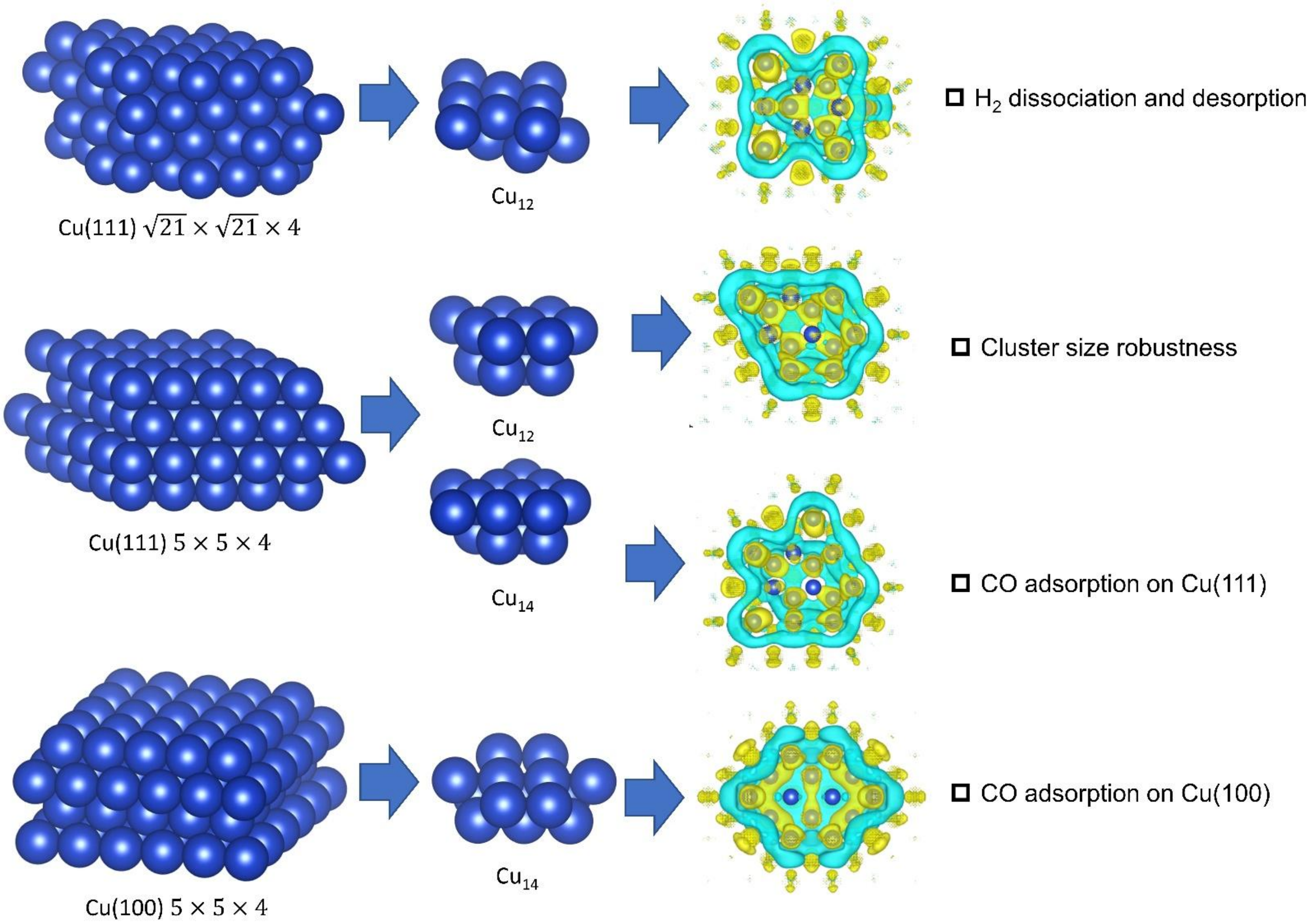

Fig. S1 Embedded Cu cluster models and representative optimized embedding potentials. The $Cu_{12}$ cluster used for $H_2$ dissociation/desorption was constructed from a $\sqrt{21} \times \sqrt{21} \times 4$ Cu(111) slab. The $Cu_{12}$ cluster used for the cluster-size robustness analysis and the $Cu_{14}$ cluster used for CO adsorption on Cu(111) were constructed from a $5 \times 5 \times 4$ Cu(111) slab model. The $Cu_{14}$ cluster used for CO adsorption on Cu(100) was constructed from a $5 \times 5 \times 4$ Cu(100) slab model. Representative optimized embedding potentials are shown as isosurfaces for each cluster model.

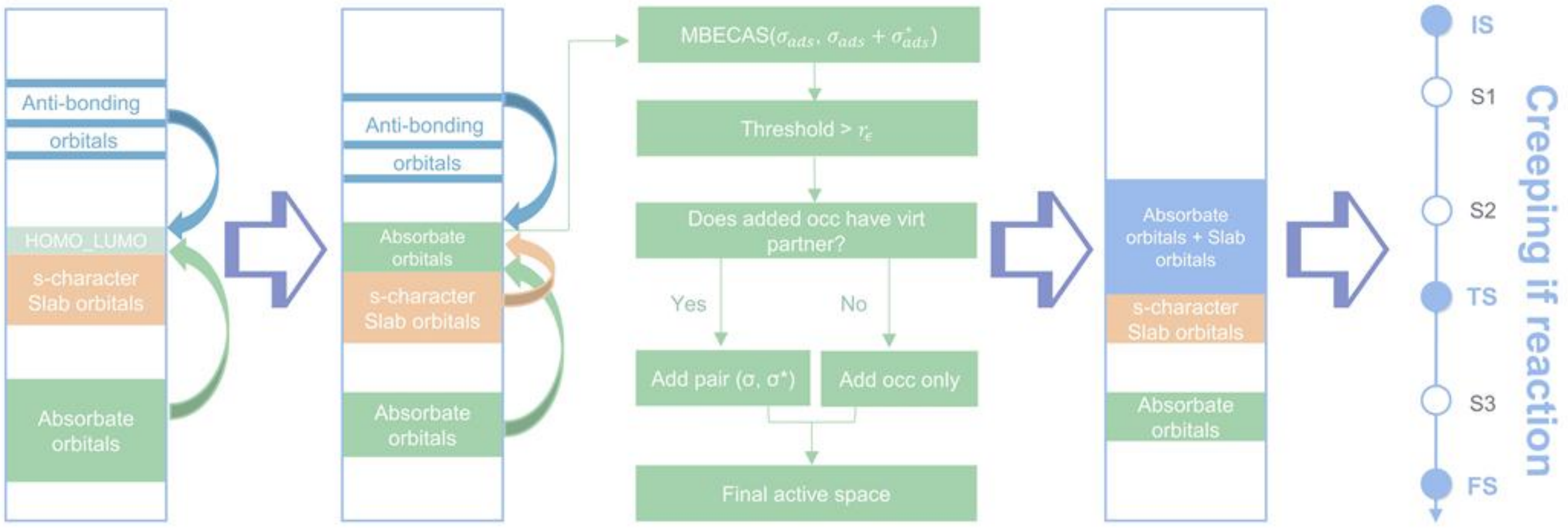

**Fig. S2 Schematic illustration of the MBECAS-SR active-space construction scheme.**

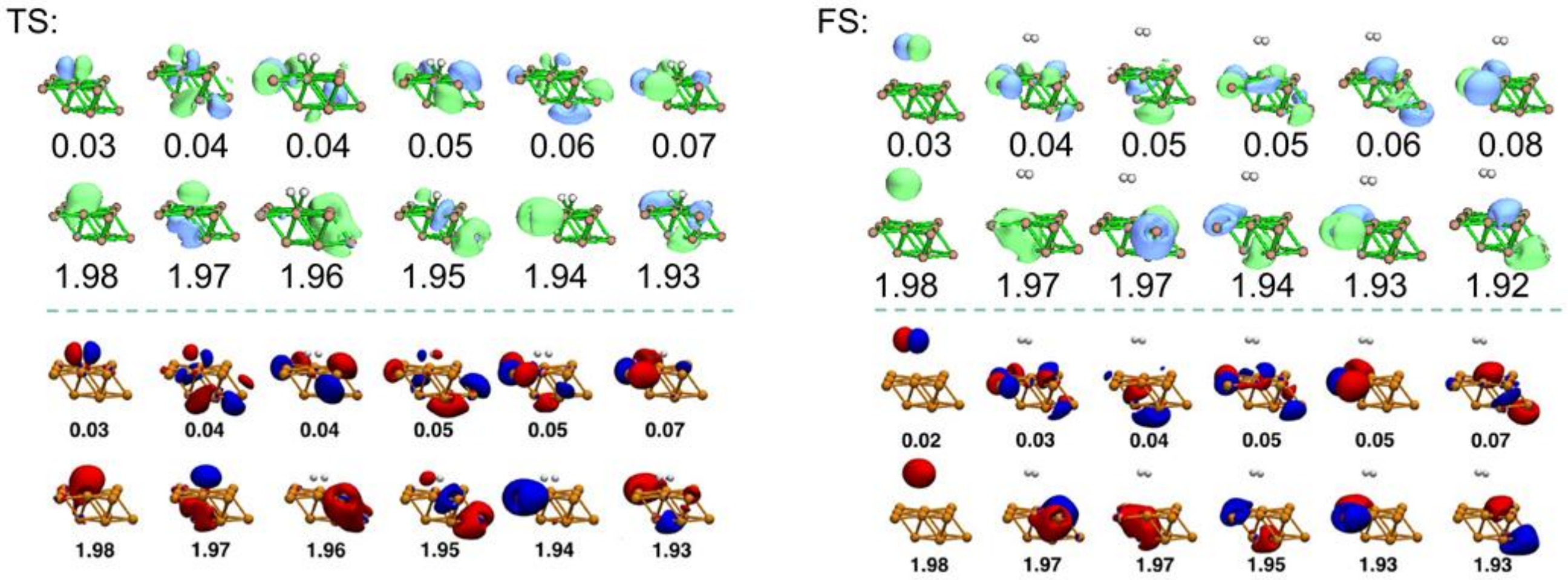

**Fig. S3 Active-orbital comparison for the TS and FS geometries of $H_2$ dissociation/desorption on Cu (111).** For each geometry, orbitals above the dashed line are selected by the present MBECAS-SR protocol, whereas orbitals below the dashed line are the manually selected active orbitals from the previous DFET-based ENEVPT2 study[11]. Natural orbital occupations are shown below the corresponding orbitals. All orbitals were visualized using Multiwfn[32,33] with an isovalue of 0.04.

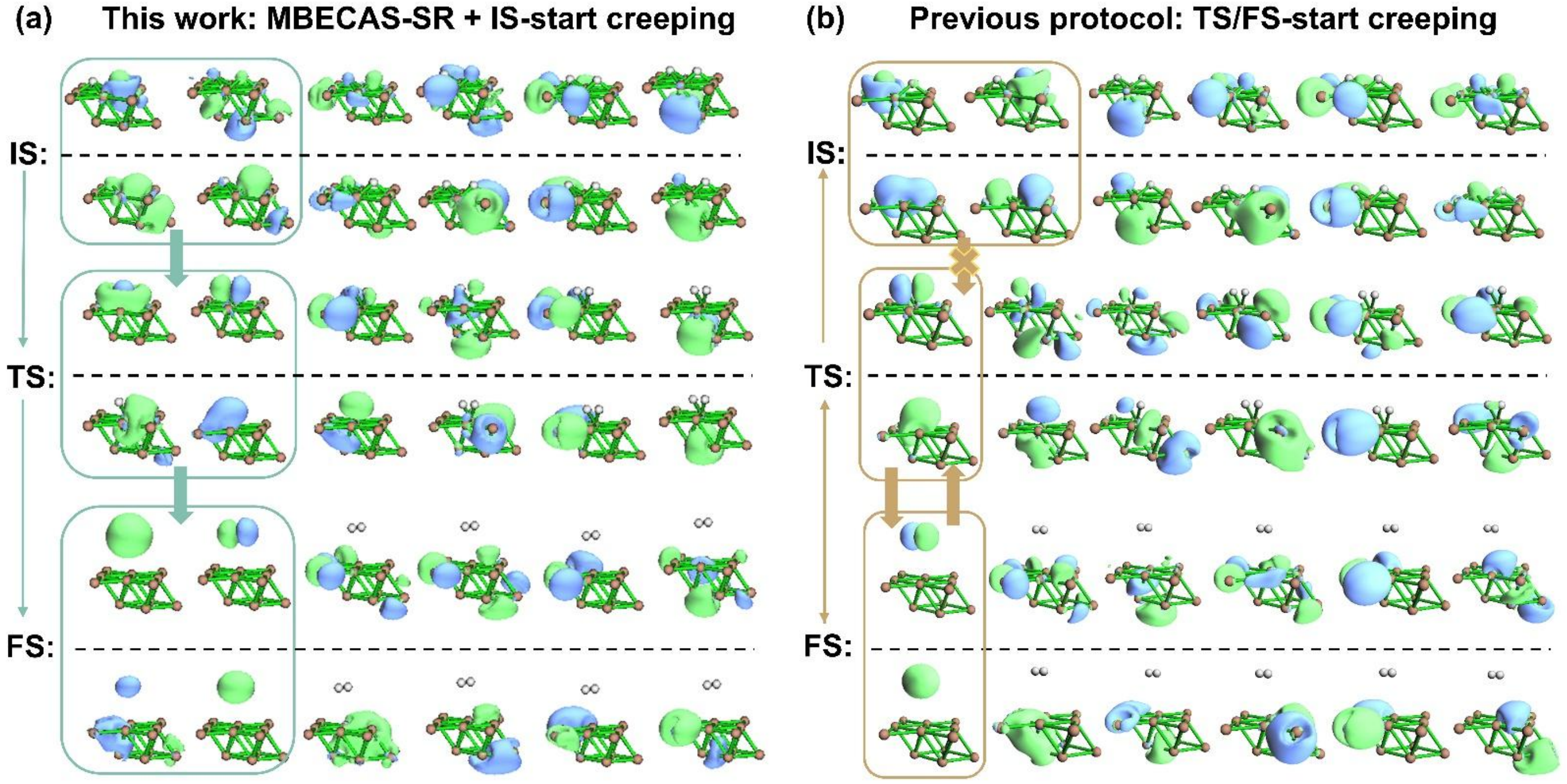

**Fig. S4 Comparison of active orbitals for $H_2$ dissociation/desorption on Cu (111) obtained from different active-space construction and orbital-propagation protocols.** (a) Active orbitals obtained in the present work using MBECAS-SR combined with creeping initiated from the fully dissociated IS. (b) Active orbitals reproduced following the previous DFET-based ENEVPT2 study[11], where creeping was initiated from the IS or TS. The rows correspond to the IS, TS, and FS geometries, and the arrows indicate the direction of orbital propagation. For each geometry, the orbitals above the dashed line are virtual active orbitals, while those below the dashed line are occupied active orbitals. To facilitate a direct comparison of orbital consistency across geometries and propagation protocols, the orbital order has been rearranged. The highlighted boxes mark the adsorbate-centered orbital pairs discussed in the text. All orbitals are visualized with an isovalue of 0.04.

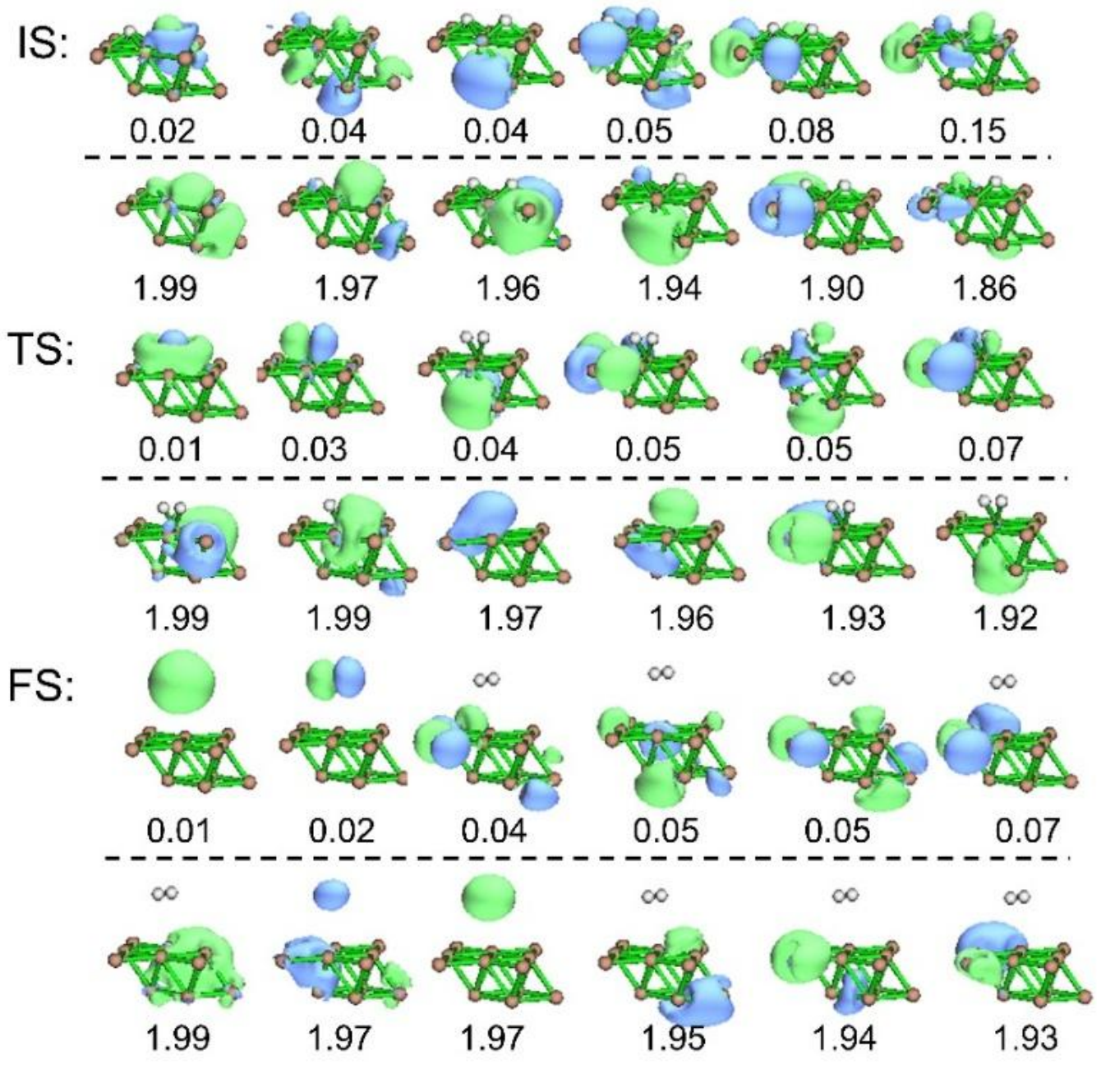

**Fig. S5 Selected natural orbitals for the $H_2$ dissociation/desorption on Cu (111).** The three panels correspond to the IS, TS, and FS geometries, respectively. For each geometry, orbitals above the dashed line

are unoccupied/virtual active orbitals, whereas orbitals below the dashed line are occupied active orbitals. The numbers below the orbital plots denote the corresponding natural orbital occupations. The orbital isosurfaces are plotted at an isovalue of 0.04.

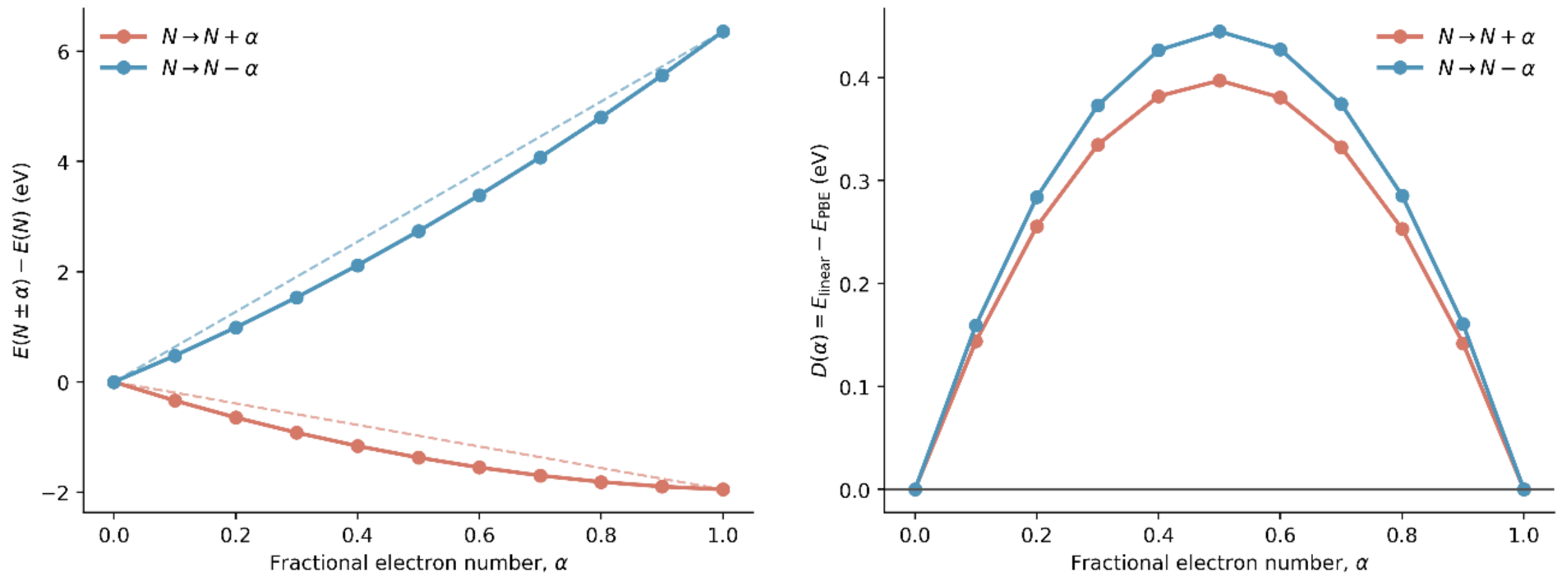

**Fig. S6 Fractional-electron energetic delocalization-error analysis for the emb-DFT Hamiltonian at the $H_2$/Cu(111) transition-state geometry.** (a) Fractional-electron energy changes for electron addition (N→N+α) and removal (N→N−α) relative to the neutral N-electron state. Solid lines show the calculated fractional-electron energies, whereas dashed lines show the piecewise-linear reference between adjacent integer-electron endpoints defined in Section 3.2. (b) Corresponding deviation from piecewise linearity, D±(α), as defined in Section 3.2. Positive values indicate artificial stabilization of fractional-charge states and hence an energetic delocalization error. The maximum deviations occur near α = 0.5 and are approximately 0.40 eV for electron addition and 0.45 eV for electron removal. The embedding potential is kept fixed throughout the fractional-electron calculations.

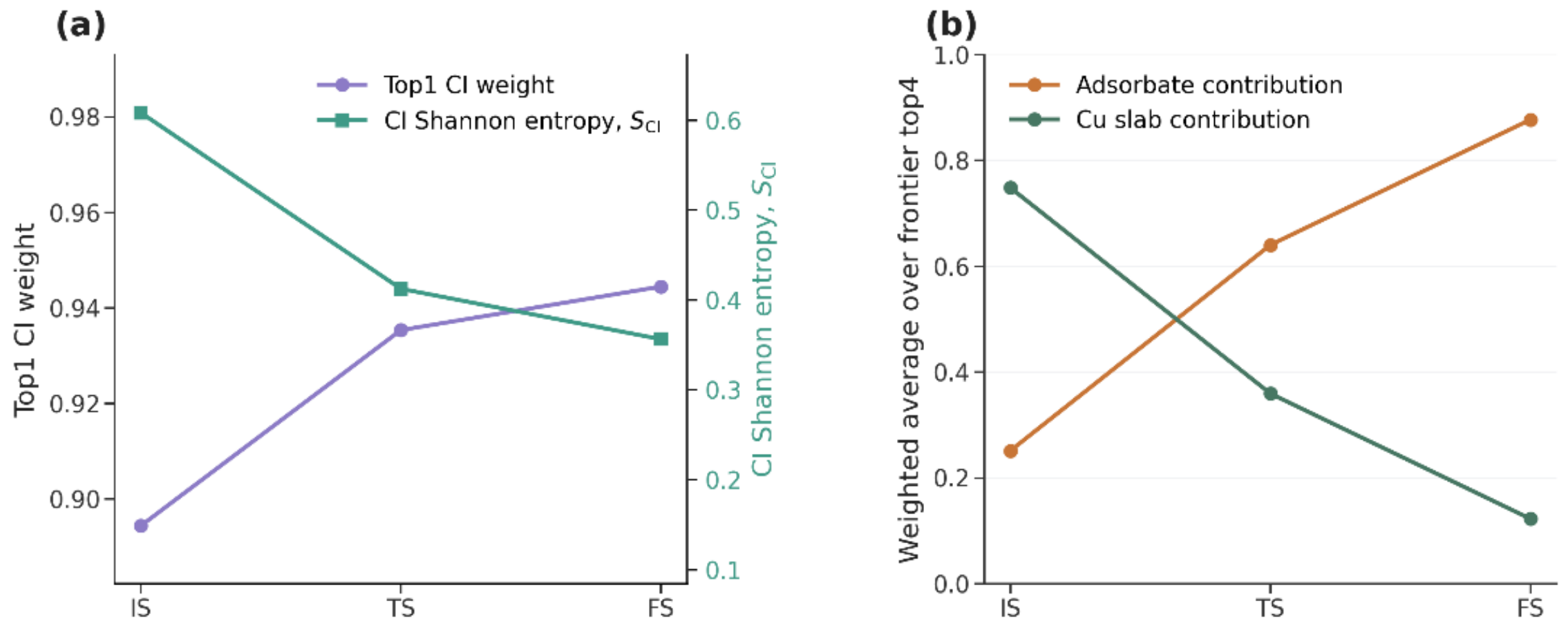

**Fig. S7 Wavefunction analysis of the H2 dissociation/desorption along the reaction coordinates on the Cu (111) surface.** (a) CI-based wavefunction-complexity diagnostics. A smaller top-1 CI weight and a larger $S_{\mathrm{CI}}$ indicate a more distributed CI expansion and stronger multiconfigurational character. (b) Evolution of frontier-orbital descriptors. The adsorbate and Cu-slab contributions are evaluated as weighted averages over the top four frontier active orbitals selected at each IS, TS, and FS geometry.

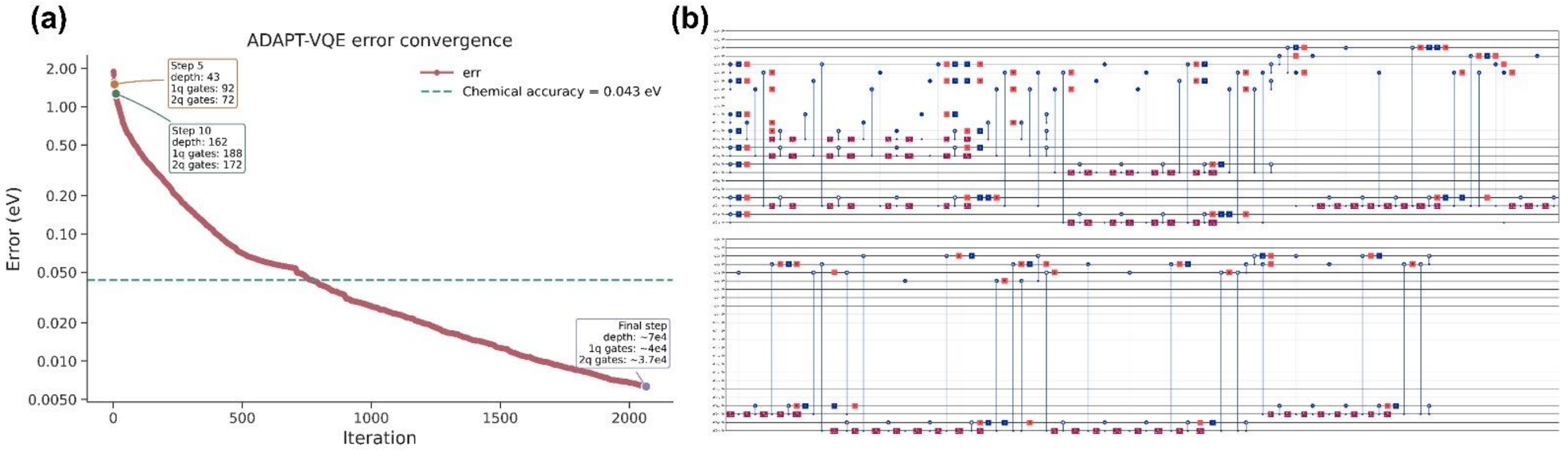

**Fig. S8 ADAPT-UCCGSD error convergence and representative quantum circuit for the IS of $H_2$ dissociation/desorption on Cu (111).** (a) ADAPT-VQE energy-error convergence as a function of iteration. The dashed horizontal line denotes the chemical-accuracy threshold of 0.043 eV. The labeled boxes indicate representative ADAPT steps used for circuit construction, with the corresponding circuit depth and numbers of one- and two-qubit gates. (b) Quantum circuit corresponding to the selected ADAPT-UCCGSD ansatz in step-10 for the IS.

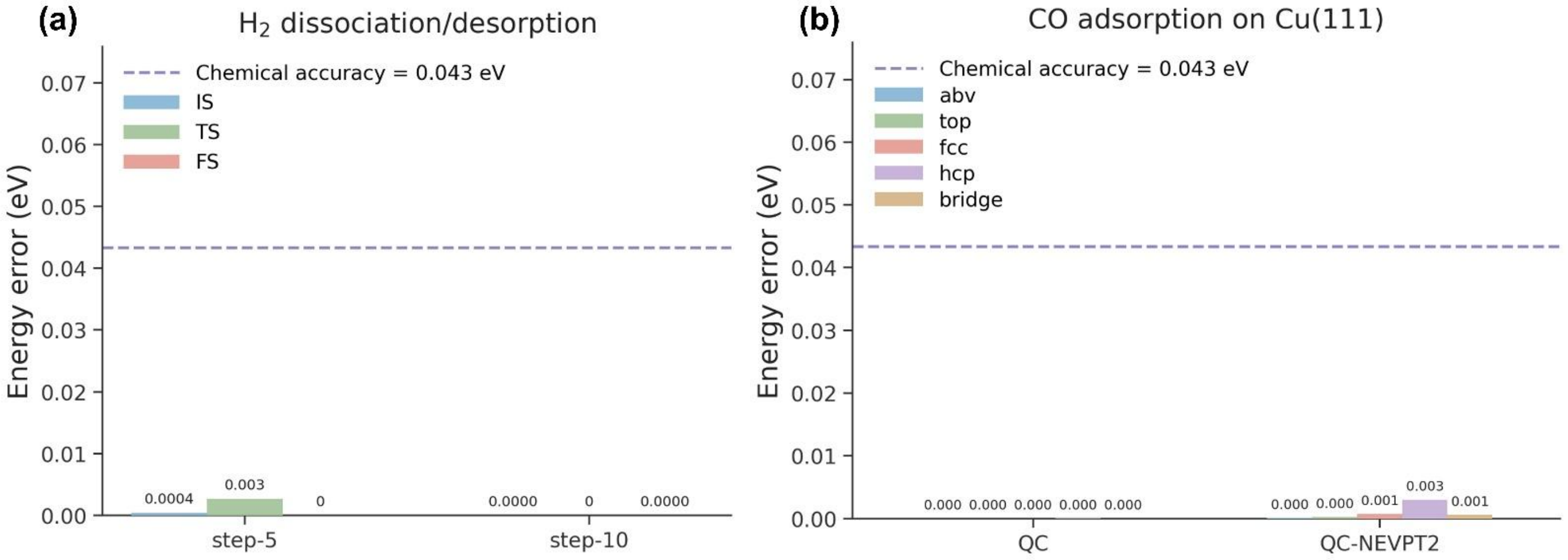

**Fig. S9 Hardware-sampled QC energy errors for $H_2$ dissociation/desorption and CO adsorption on Cu (111).** (a) Energy errors for the IS, TS, and FS of $H_2$ dissociation/desorption obtained from the step-5 and step-10 hardware-executed ADAPT circuits after configuration recovery and QSCI diagonalization. (b) Energy errors for CO adsorption configurations on Cu (111), including the above-surface (abv), top, fcc, hcp, and bridge configurations. For CO adsorption, the hardware data were obtained from the step-10 hardware-executed ADAPT circuits followed by configuration recovery and QSCI diagonalization, and the corresponding QSCI and QSCI + NEVPT2 energy errors are shown. The dashed horizontal lines denote the chemical-accuracy threshold of 0.043 eV.

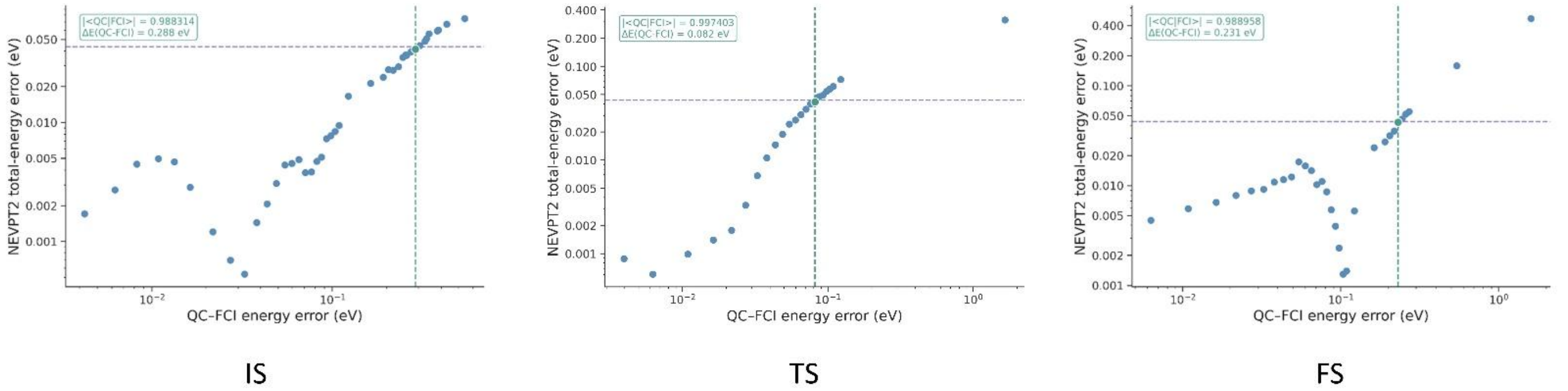

**Fig. S10 NEVPT2-assisted error reduction for intermediate quantum-computing states in $H_2$ dissociation/desorption on Cu(111).** The sampled states were taken from the ADAPT-UCCGSD convergence trajectories for the IS, TS, and FS structures. For each intermediate state, the wave-function overlap with the FCI reference and the corresponding QC and QC + NEVPT2 energy errors are shown, illustrating how SC-NEVPT2 reduces the residual energy error as the quality of the active-space state improves.

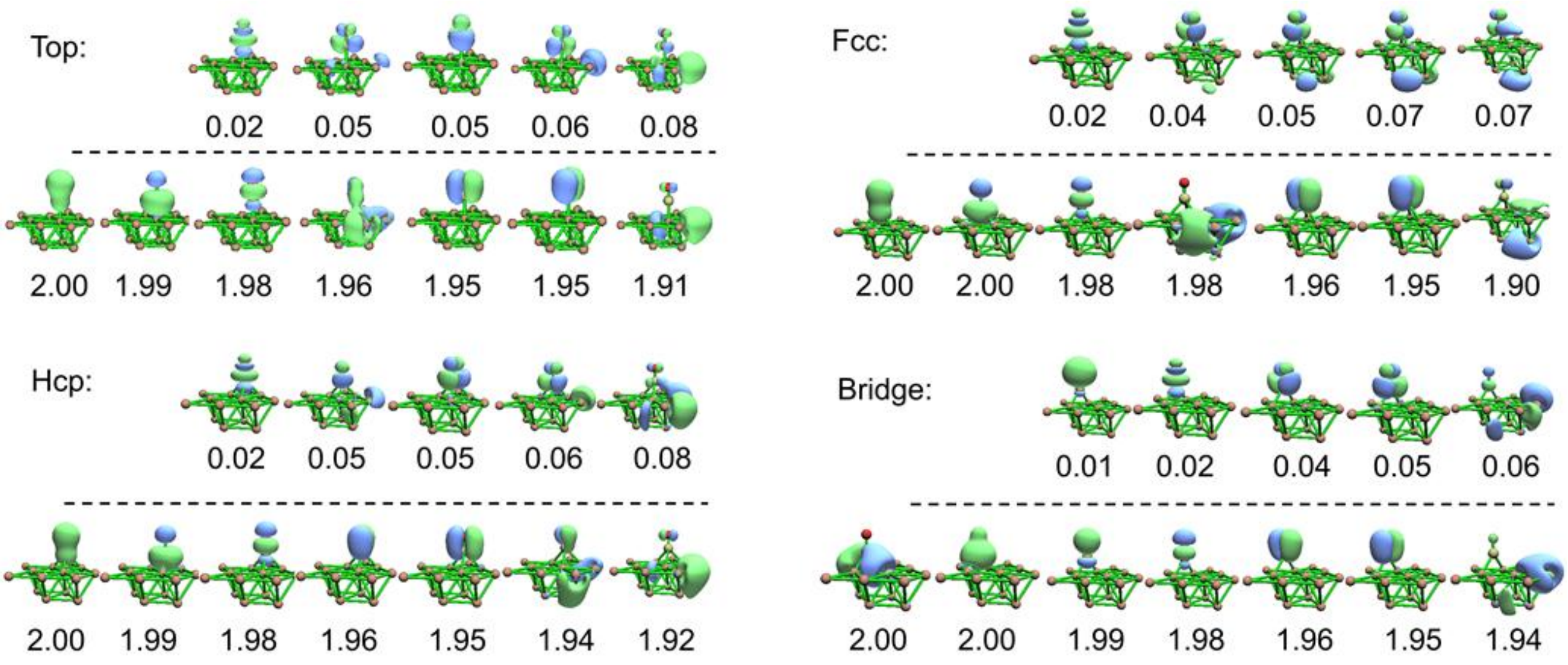

**Fig. S11 Selected active orbitals for CO adsorption on Cu (111).** The active space contains 14 electrons in 12 orbitals, CAS (14e,12o). Orbitals above the dashed line are unoccupied/virtual active orbitals, whereas orbitals below the dashed line are occupied active orbitals. The numbers below the orbital plots denote the corresponding natural orbital occupations. The orbital isosurfaces are plotted at an isovalue of 0.04.

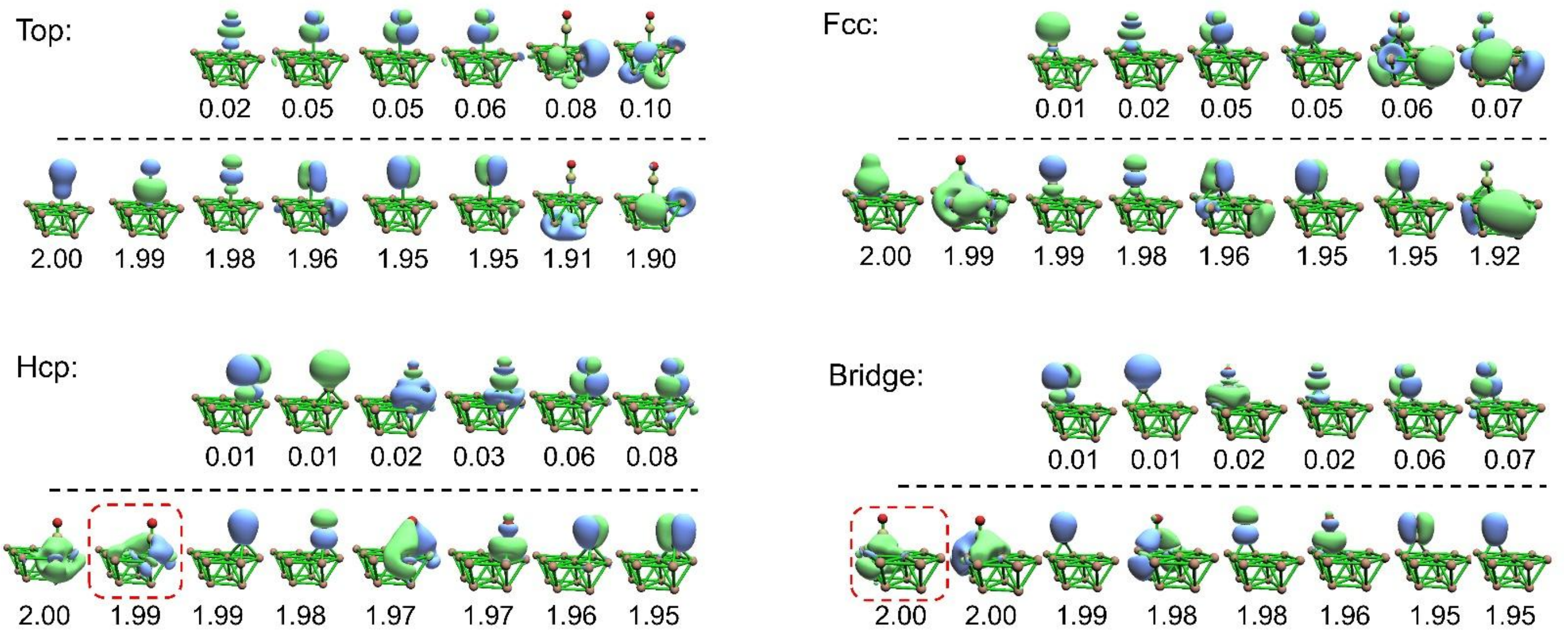

**Fig. S12 Selected active orbitals for cluster size robustness analysis in CO adsorption puzzle on Cu (111).** The active space contains 16 electrons in 14 orbitals, CAS (16e,14o). Orbitals above the dashed line are unoccupied/virtual active orbitals, whereas orbitals below the dashed line are occupied active orbitals. The red dotted boxes highlight active orbitals with pronounced boundary-Cu 3d character. The numbers below the orbital plots denote the corresponding natural orbital occupations. The orbital isosurfaces are plotted at an isovalue of 0.04.

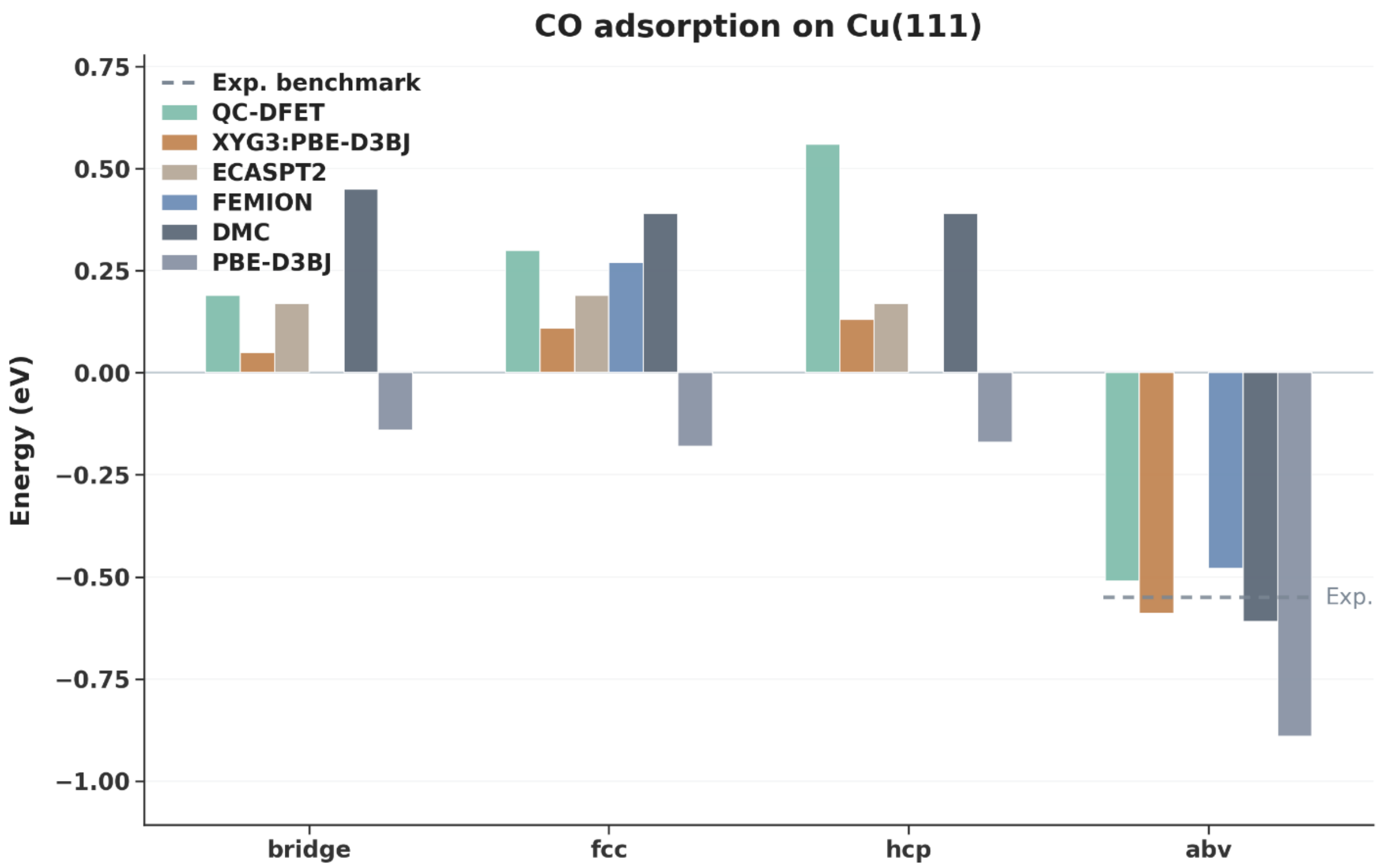

**Fig. S13 Site-resolved energetics of CO adsorption on Cu (111).** Energies are shown relative to the top adsorption site for the bridge, fcc, hcp, and abv configurations. Colored bars denote QC-DFET, XYG3: PBE-D3BJ, ECASPT2, FEMION, DMC, and PBE-D3BJ results, as indicated in the legend. The dashed horizontal line marks the experimental benchmark for the abv–top energy difference. Positive values indicate configurations higher in energy than the top site, whereas negative values indicate configurations lower in energy than the top site.

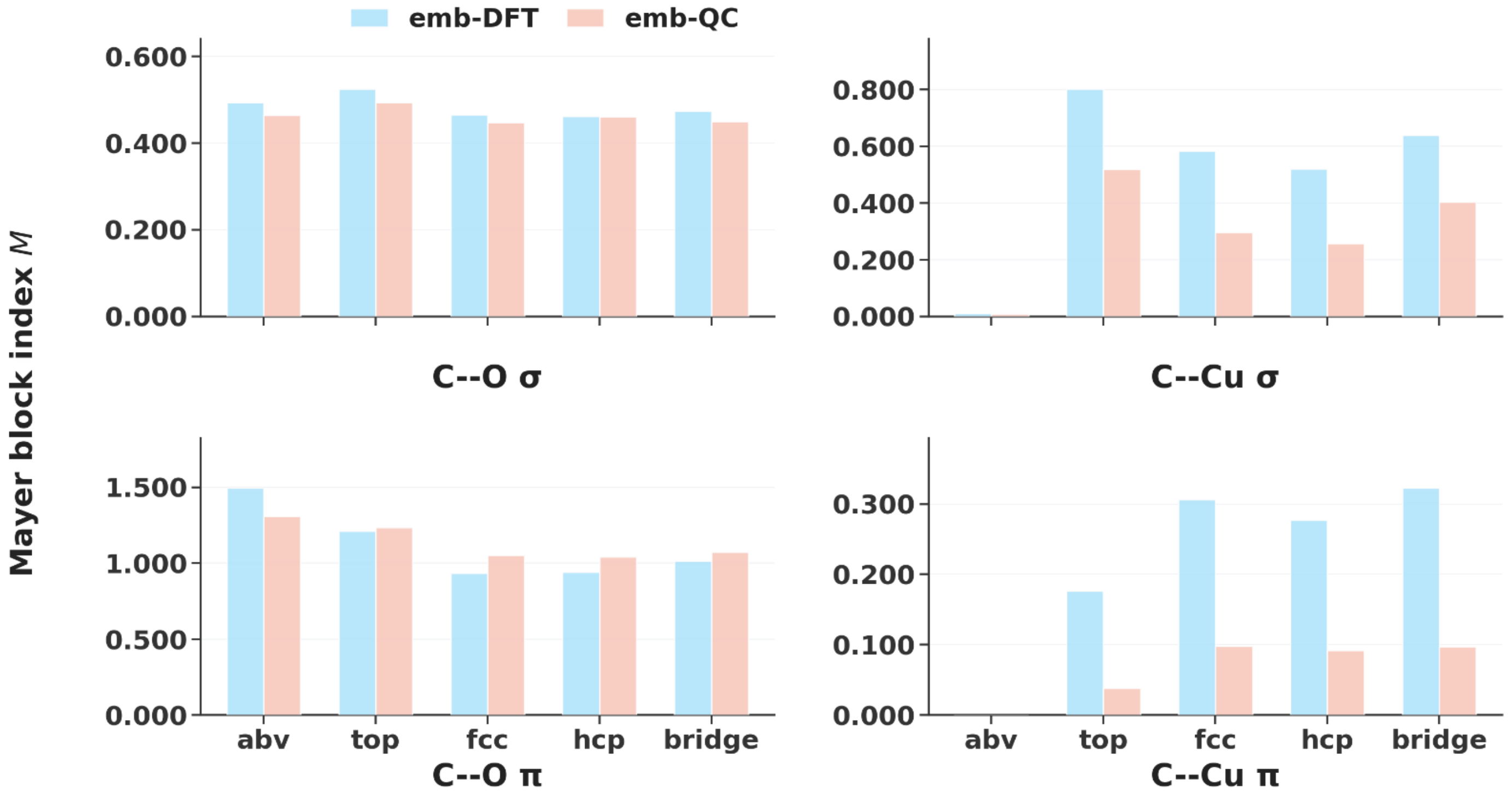

**Fig. S14 Absolute orbital-resolved Mayer block indices for CO adsorption on Cu(111).** The generalized Mayer block index M is decomposed into four orbital channels: the internal C–O σ component, the C–Cu σ coupling, the internal C–O π component, and the C–Cu π coupling. Results are shown for the abv, top, fcc, hcp, and bridge configurations. Blue bars denote emb-DFT results, and peach bars denote emb-QC results.

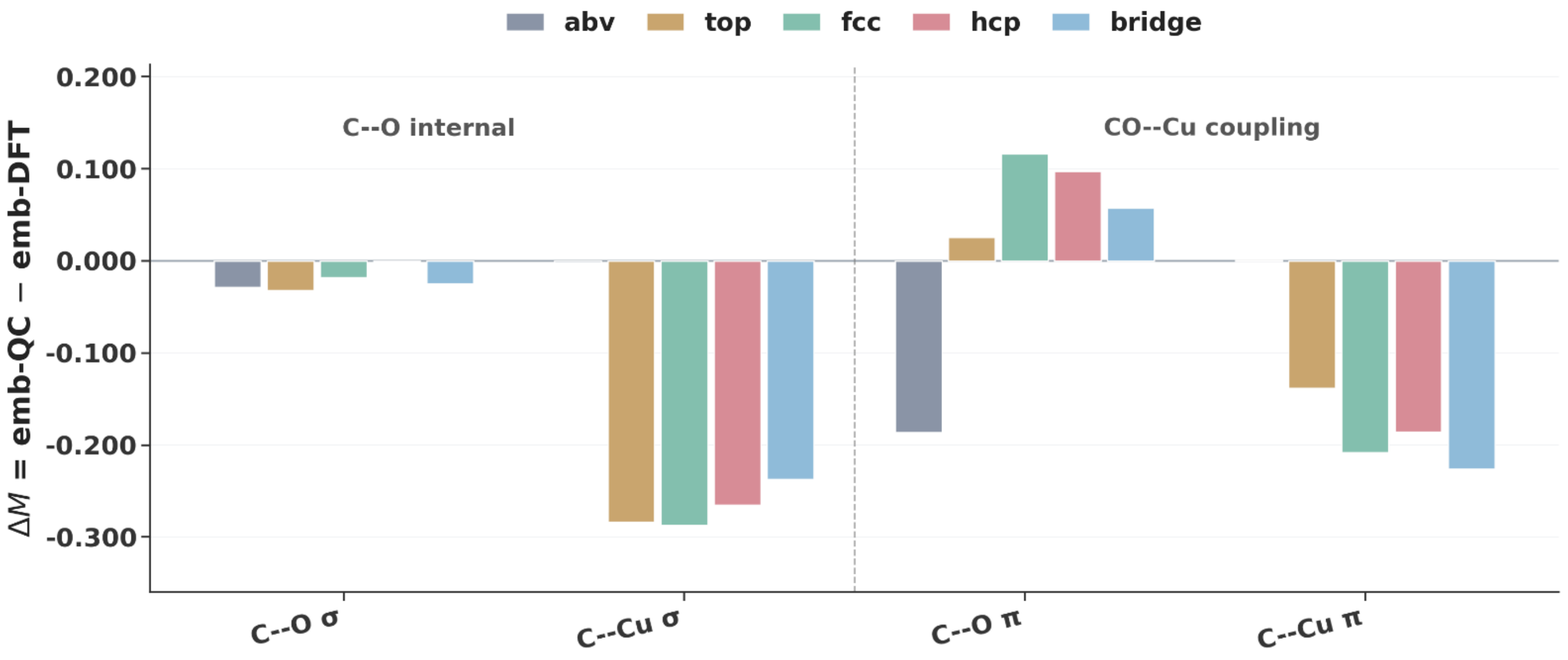

**Fig. S15 Orbital-resolved Mayer bond-order correction for CO adsorption on Cu(111).** The differences in generalized Mayer block indices are defined as ΔM = M(emb-QC) − M(emb-DFT). Results are shown for the abv, top, fcc, hcp, and bridge configurations. The four grouped channels correspond to the internal C–O σ component, the C–Cu σ coupling, the internal C–O π component, and the C–Cu π coupling. Different bar colors denote different adsorption configurations, as indicated in the legend. The vertical dashed line separates the C–O internal channels from the CO–Cu coupling channels.

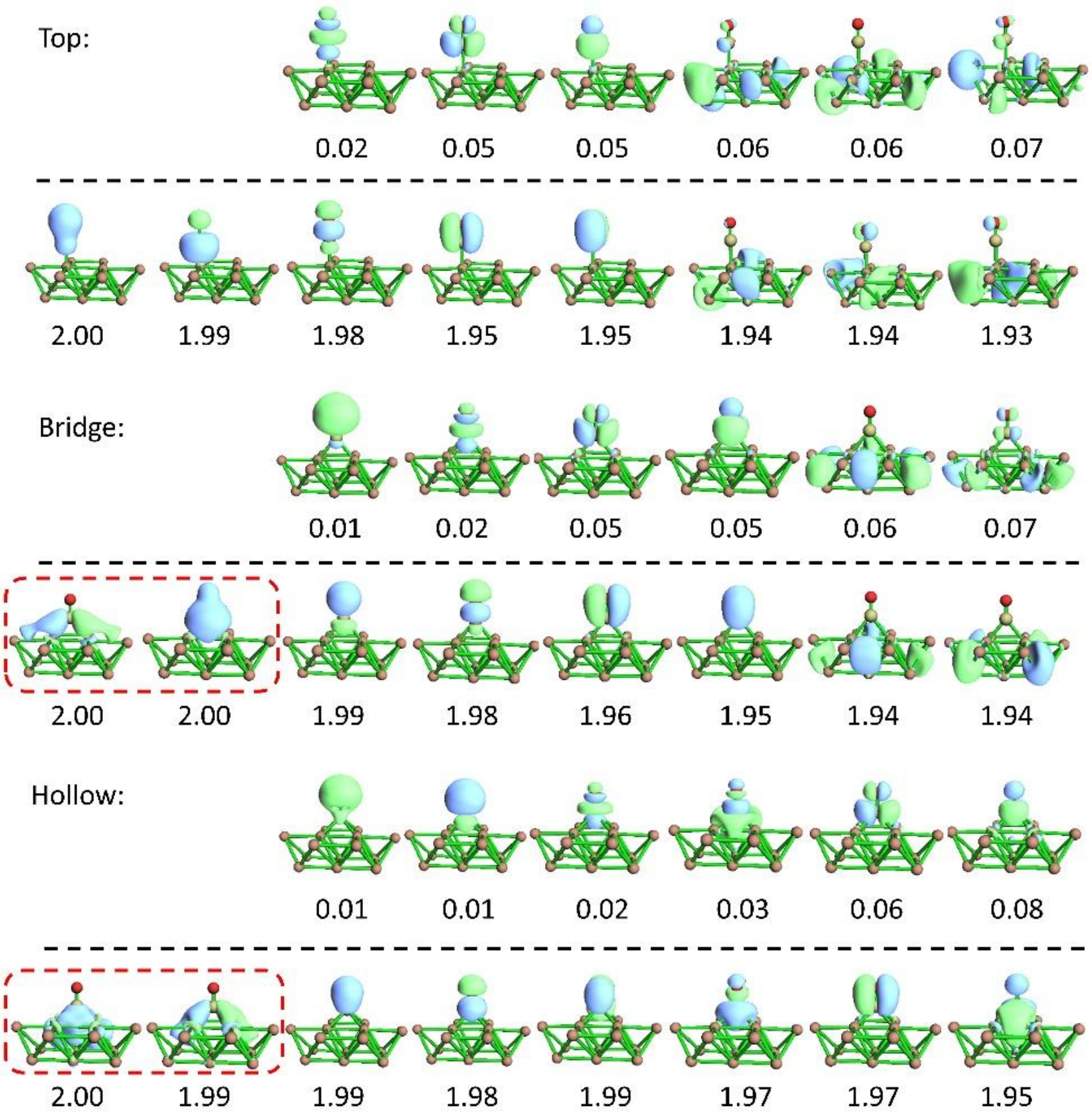

**Fig. S16 Selected natural orbitals within CAS(16e, 14o) for the CO adsorption on Cu (100).** Orbitals above the dashed line are unoccupied/virtual active orbitals, whereas orbitals below the dashed line are occupied active orbitals. The red dotted boxes highlight active orbitals with pronounced boundary-Cu 3d character. The numbers below the orbital plots denote the corresponding natural orbital occupations. The orbital isosurfaces are plotted at an isovalue of 0.04.

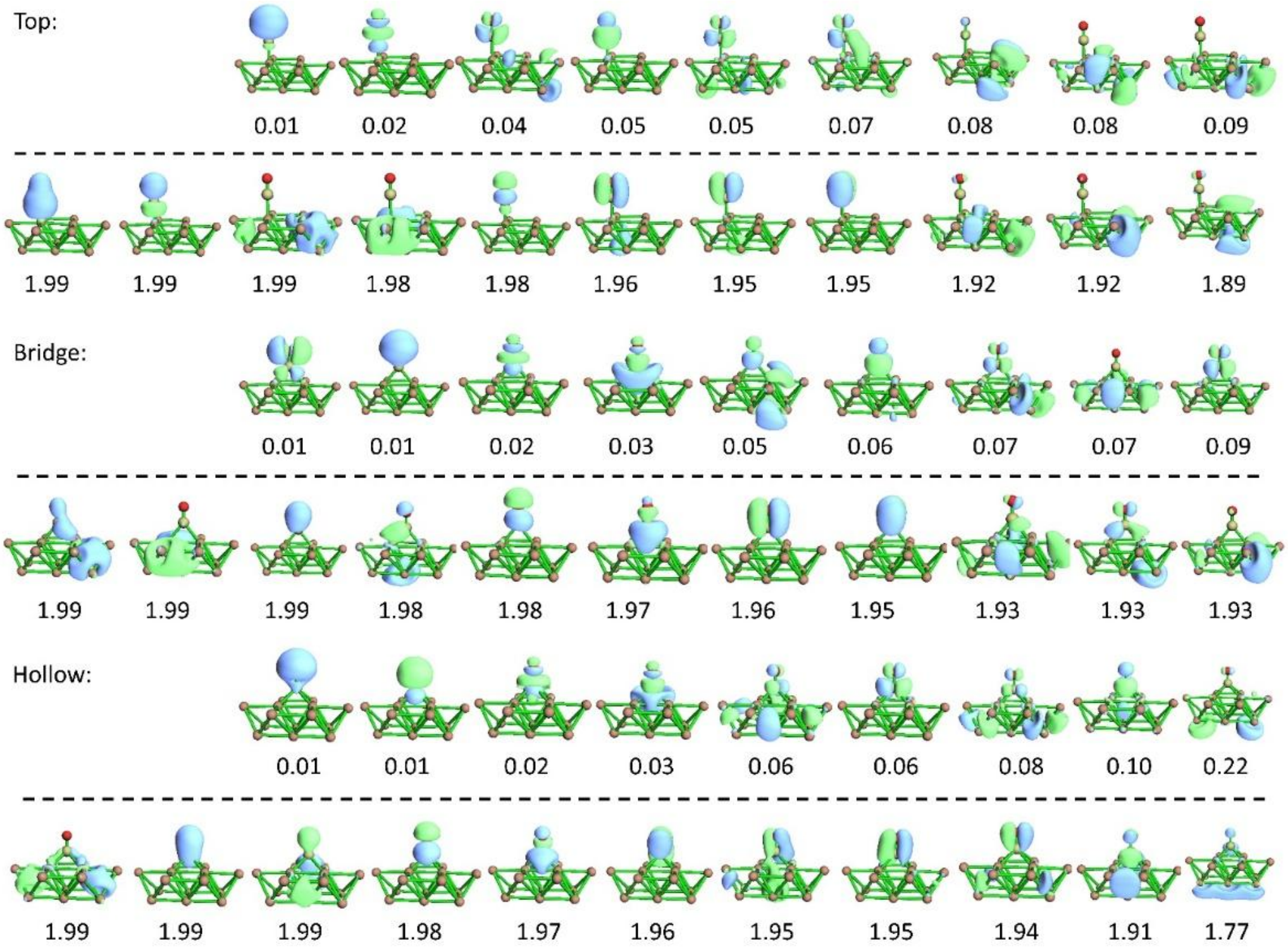

**Fig. S17 Selected natural orbitals within CAS(22e, 20o) for the CO adsorption on Cu (100).** Orbitals above the dashed line are unoccupied/virtual active orbitals, whereas orbitals below the dashed line are occupied active orbitals. The numbers below the orbital plots denote the corresponding natural orbital occupations. The orbital isosurfaces are plotted at an isovalue of 0.04.

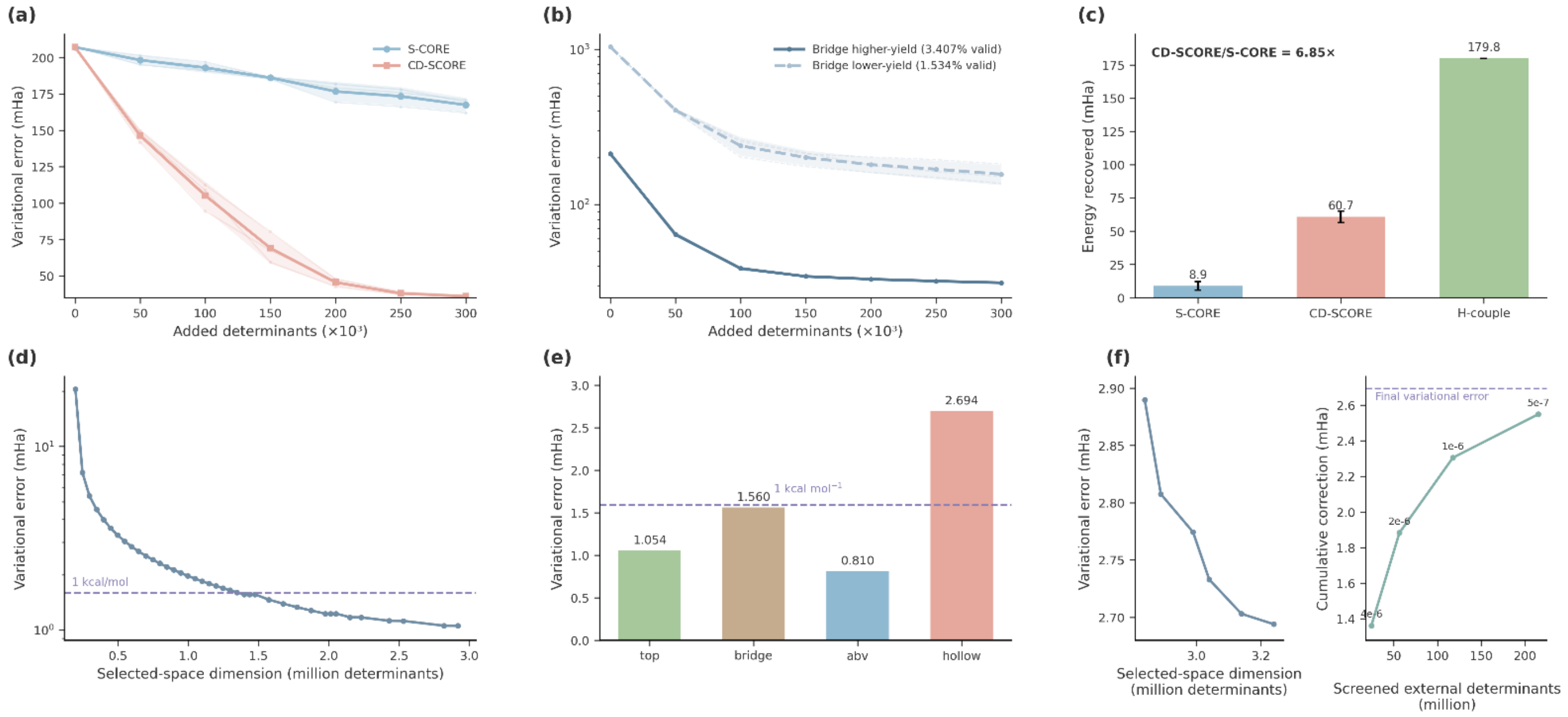

**Fig. S18. Configuration-recovery benchmarks and variational convergence for the 40-qubit CO/Cu(100) calculations.** (a) S-CORE and CD-SCORE starting from the same measured determinant subspace, with six rounds of 50,000 newly added determinants; curves and shaded regions show means and standard deviations over three seeds. (b) CD-SCORE applied to two independently measured quantum datasets at the bridge site, with higher-yield (3.407% valid) and lower-yield (1.534% valid) samples, each following the same six-round expansion budget; the vertical axis is logarithmic. (c) Equal-budget single-round comparison of S-CORE, CD-SCORE, and H-couple from a common initial subspace, with 50,000 added determinants per strategy; stochastic results include seed variability. (d) Convergence of the recovered 40-qubit variational energy for the top site. (e) Final variational errors for top, bridge, abv, and hollow; the dashed line denotes 1 kcal mol$^{-1}$. (f) Late-stage variational convergence for hollow (left) and the cumulative screened Epstein–Nesbet second-order estimate for determinants outside the selected subspace within the same 40-qubit configuration space (right).

## 7. Supplementary Tables

**Table. S1** CO adsorption energetics on the $Cu_{12}$ embedded cluster for Cu(111) using CAS(16e,14o). Energies are reported relative to the top adsorption site.

| Adsorption site | Energy relative to top / eV |
|---|---|
| top | 0.00 |
| bridge | 0.51 |
| fcc | 0.46 |
| hcp | 0.73 |
| abv | 0.54 |

**Table. S2** CO adsorption energetics on Cu(100) from the CAS(16e,14o) QC-DFET calculation. Energies are reported relative to the top adsorption site; the abv–top separation corresponds to the magnitude of the top-site adsorption energy.

| Adsorption site | Energy relative to top / eV |
|---|---|
| top | 0.00 |
| abv | 0.56 |
| bridge | 0.61 |
| hollow | 0.98 |

**Table S3** Variational reference-space errors and EN-SC-NEVPT2 correlation corrections for the 40-qubit CO/Cu(100) calculations.

| Adsorption site (CSF space) | QSCI error (mHa) | Full-CFG error (mHa) | Selected-CSF error (mHa) | EN-SC-NEVPT2 (Ha) |
|---|---|---|---|---|
| Top (1.0M) | 1.054 | 0.633 | 1.120 | −11.566304 |
| ABV (1.0M) | 0.810 | 0.517 | 0.528 | −11.526729 |
| Bridge (1.0M) | 1.560 | 1.079 | 1.195 | −11.577718 |
| Hollow (1.0M) | 2.694 | 1.999 | 3.269 | −11.574912 |

Note. The first three numerical columns are variational errors relative to the same-site SHCI reference. The final column gives the EN-SC-NEVPT2 second-order correlation energy, not an error. "Full-CFG" includes all singlet CSFs generated from the QSCI-derived CFG pool; all selected-CSF references use approximately 1M CSFs.

**References:**

1. Kresse, G. & Furthmüller, J. Efficiency of ab-initio total energy calculations for metals and semiconductors using a plane-wave basis set. Computational Materials Science 6, 15–50 (1996).
2. Kresse, G. & Furthmüller, J. Efficient iterative schemes for ab initio total-energy calculations using a plane-wave basis set. Physical Review B 54, 11169 (1996).

3. Blöchl, P. E. Projector augmented-wave method. Physical Review B 50, 17953 (1994).
4. Perdew, J. P., Burke, K. & Ernzerhof, M. Generalized gradient approximation made simple. Physical Review Letters 77, 3865 (1996).
5. Grimme, S., Antony, J., Ehrlich, S. & Krieg, H. A consistent and accurate ab initio parametrization of density functional dispersion correction (DFT-D) for the 94 elements H–Pu. The Journal of Chemical Physics 132, 154104 (2010).
6. Grimme, S., Ehrlich, S. & Goerigk, L. Effect of the damping function in dispersion corrected density functional theory. Journal of Computational Chemistry 32, 1456–1465 (2011).
7. Becke, A. D. & Johnson, E. R. A density-functional model of the dispersion interaction. The Journal of Chemical Physics 123, 154101 (2005).
8. Monkhorst, H. J. & Pack, J. D. Special points for Brillouin-zone integrations. Physical Review B 13, 5188 (1976).
9. Methfessel, M. & Paxton, A. T. High-precision sampling for Brillouin-zone integration in metals. Physical Review B 40, 3616 (1989).
10. Henkelman, G., Uberuaga, B. P. & Jónsson, H. A climbing image nudged elastic band method for finding saddle points and minimum energy paths. The Journal of Chemical Physics 113, 9901–9904 (2000).
11. Zhao, Q., Zhang, X., Martirez, J. M. P. & Carter, E. A. Benchmarking an embedded adaptive sampling configuration interaction method for surface reactions: $H_2$ desorption from and $CH_4$ dissociation on Cu (111). Journal of Chemical Theory and Computation 16, 7078–7088 (2020).
12. Yu, K. & Carter, E. A. VASPEmbedding. GitHub, https://github.com/EACcodes/VASPEmbedding.
13. Krauter, C. M. & Carter, E. A. EmbeddingIntegralGenerator. GitHub, https://github.com/EACcodes/EmbeddingIntegralGenerator.
14. Sun, Q. et al. PySCF: the Python-based simulations of chemistry framework. Wiley Interdisciplinary Reviews: Computational Molecular Science 8, e1340 (2018).
15. Sun, Q. et al. Recent developments in the PySCF program package. The Journal of Chemical Physics 153, 024109 (2020).
16. Smith, J. E. T. et al. Cheap and near exact CASSCF with large active spaces. The Journal of Chemical Theory and Computation 13, 5468–5478 (2017).
17. Holmes, A. A., Tubman, N. M. & Umrigar, C. J. Heat-bath configuration interaction: An efficient selected configuration interaction algorithm inspired by heat-bath sampling. The Journal of Chemical Theory and Computation 12, 3674–3680 (2016).
18. Ugandi, M. & Roemelt, M. A configuration-based heatbath-CI for spin-adapted multireference electronic structure calculations with large active spaces. Journal of Computational Chemistry 44, 2374–2390 (2023).
19. Ugandi, M. & Roemelt, M. An Uncontracted Epstein–Nesbet Perturbation Theory Approximation to SC-NEVPT2 Based on Spin-Pure Selected CI Wave Functions. Journal of Chemical Theory and Computation 22, 7164–7176 (2026).
20. Dunning, T. H. Jr. Gaussian basis sets for use in correlated molecular calculations. I. The atoms boron through neon and hydrogen. The Journal of Chemical Physics 90, 1007–1023 (1989).
21. Peterson, K. A. & Puzzarini, C. Systematically convergent basis sets for transition metals. II. Pseudopotential-based correlation consistent basis sets for the group 11 (Cu, Ag, Au) and 12 (Zn, Cd, Hg) elements. Theoretical Chemistry Accounts 114, 283–296 (2005).
22. Dolg, M., Wedig, U., Stoll, H. & Preuss, H. Energy-adjusted ab initio pseudopotentials for the first row transition elements. The Journal of Chemical Physics 86, 866–872 (1987).
23. Zeng, X. et al. Accurate chemical reaction modeling on a noisy intermediate-scale quantum computer with an active space-based workflow. The Journal of Physical Chemistry Letters 16, 11709–11716 (2025).
24. Delcey, M. G., Pedersen, T. B., Aquilante, F. & Lindh, R. Analytical gradients of the state-average

complete active space self-consistent field method with density fitting. The Journal of Chemical Physics 143, 044110 (2015).
25. Aquilante, F. et al. Accurate ab initio density fitting for multiconfigurational self-consistent field methods. The Journal of Chemical Physics 129, 024113 (2008).
26. Stoychev, G. L., Auer, A. A. & Neese, F. Automatic generation of auxiliary basis sets. Journal of Chemical Theory and Computation 13, 554–562 (2017).
27. Cao, C. et al. Quantum many-body simulations of catalytic metal surfaces. arXiv preprint arXiv:2508.13036 (2025).
28. Fan, Y. et al. $Q^2$ Chemistry: a quantum computation platform for quantum chemistry. arXiv preprint arXiv:2208.10978 (2022).
29. Saki, A. A. et al. Qiskit addon: sample-based quantum diagonalization. GitHub, https://github.com/Qiskit/qiskit-addon-sqd (2024).
30. Zhang, H., Zeng, X., Li, Z. & Zhou, Y. Hamiltonian-Guided Autoregressive Selected-Configuration Interaction Achieves Chemical Accuracy in Strongly Correlated Systems. Journal of Chemical Theory and Computation 21, 12622–12633 (2025).
31. Robledo-Moreno, J. et al. Chemistry beyond the scale of exact diagonalization on a quantum-centric supercomputer. Science Advances 11, eadu9991 (2025).
32. Lu, T. & Chen, F. Multiwfn: a multifunctional wavefunction analyzer. Journal of Computational Chemistry 33, 580–592 (2012).
33. Lu, T. A comprehensive electron wavefunction analysis toolbox for chemists, Multiwfn. The Journal of Chemical Physics 161, 082503 (2024).